\documentclass[floatfix, noshowpacs, preprintnumbers, twocolumn, amsmath, amssymb, aps, superscriptaddress, prb]{revtex4}

\pdfoutput=1

\usepackage{graphicx, xcolor}
\usepackage[caption=false]{subfig}
\usepackage{soul}
\usepackage{mathrsfs}
\usepackage{float}

\newcommand{\bra}[1]{\left< #1 \right|} 
\newcommand{\ket}[1]{\left| #1 \right>}

\begin{document}

\title{Quantum Computation for Pricing the Collateralized Debt Obligations}

\author{Hao Tang}
\email{htang2015@sjtu.edu.cn}
\affiliation{Center for Integrated Quantum Information Technologies (IQIT), School of Physics and Astronomy and State Key Laboratory of Advanced Optical Communication Systems and Networks, Shanghai Jiao Tong University, Shanghai 200240, China}
\affiliation{CAS Center for Excellence and Synergetic Innovation Center in Quantum Information and Quantum Physics, University of Science and Technology of China, Hefei, Anhui 230026, China}

\author{Anurag Pal}
\affiliation{Center for Integrated Quantum Information Technologies (IQIT), School of Physics and Astronomy and State Key Laboratory of Advanced Optical Communication Systems and Networks, Shanghai Jiao Tong University, Shanghai 200240, China}
\affiliation{CAS Center for Excellence and Synergetic Innovation Center in Quantum Information and Quantum Physics, University of Science and Technology of China, Hefei, Anhui 230026, China}

\author{Tian-Yu Wang}
\affiliation{Center for Integrated Quantum Information Technologies (IQIT), School of Physics and Astronomy and State Key Laboratory of Advanced Optical Communication Systems and Networks, Shanghai Jiao Tong University, Shanghai 200240, China}
\affiliation{CAS Center for Excellence and Synergetic Innovation Center in Quantum Information and Quantum Physics, University of Science and Technology of China, Hefei, Anhui 230026, China}

\author{Lu-Feng Qiao}
\affiliation{Center for Integrated Quantum Information Technologies (IQIT), School of Physics and Astronomy and State Key Laboratory of Advanced Optical Communication Systems and Networks, Shanghai Jiao Tong University, Shanghai 200240, China}
\affiliation{CAS Center for Excellence and Synergetic Innovation Center in Quantum Information and Quantum Physics, University of Science and Technology of China, Hefei, Anhui 230026, China}

\author{Jun Gao}
\affiliation{Center for Integrated Quantum Information Technologies (IQIT), School of Physics and Astronomy and State Key Laboratory of Advanced Optical Communication Systems and Networks, Shanghai Jiao Tong University, Shanghai 200240, China}
\affiliation{CAS Center for Excellence and Synergetic Innovation Center in Quantum Information and Quantum Physics, University of Science and Technology of China, Hefei, Anhui 230026, China}

\author{Xian-Min Jin}
\email{xianmin.jin@sjtu.edu.cn} 
\affiliation{Center for Integrated Quantum Information Technologies (IQIT), School of Physics and Astronomy and State Key Laboratory of Advanced Optical Communication Systems and Networks, Shanghai Jiao Tong University, Shanghai 200240, China}
\affiliation{CAS Center for Excellence and Synergetic Innovation Center in Quantum Information and Quantum Physics, University of Science and Technology of China, Hefei, Anhui 230026, China}

\maketitle
\textbf{Collateralized debt obligation (CDO) has been one of the most commonly used structured financial products and is intensively studied in quantitative finance. By setting the asset pool into different tranches, it effectively works out and redistributes credit risks and returns to meet the risk preferences for different tranche investors. The copula models of various kinds are normally used for pricing CDOs, and the Monte Carlo simulations are required to get their numerical solution. Here we implement two typical CDO models, the single-factor Gaussian copula model and Normal Inverse Gaussian copula model, and by applying the conditional independence approach, we manage to load each model of distribution in quantum circuits. We then apply quantum amplitude estimation as an alternative to Monte Carlo simulation for CDO pricing. We demonstrate the quantum computation results using IBM Qiskit. Our work addresses a useful task in finance instrument pricing, significantly broadening the application scope for quantum computing in finance.
}

Quantum computing for finance applications is an emerging field with quickly growing popularity. The finance industry involves various numerical and analytical tasks, $e.g.$, derivative pricing, credit rating, forex algorithm trading, and portfolio optimization, $etc.$. They all demand heavy quantitative work, and the improved calculation speed and precision would bring significant social value. Quantum computing aims at these very targets\cite{Orus2019}. Early studies focused on improving finance models with basic quantum mechanics\cite{Baaquie2007, Zhang2010, Meng2016}. Schrodinger equations and Feynman's path integral were suggested to solve stochastic differential equations for pricing interest rate derivatives\cite{Baaquie2007}, and Heisenberg uncertainty principle was used to interpret the leptokurtic and fat-tailed distribution of stock price volatilities\cite{Meng2016}. Recent studies tend to utilize quantum advantages as a faster computing machine. Algorithms that can be implemented in quantum circuits, such as amplitude estimation\cite{Brassard2002}, quantum principle component analysis (PCA)\cite{Lloyd2014}, quantum generative adversarial network (QGAN)\cite{Lloyd2018}, the quantum-classical hybrid variational quantum eigensolver (VQE)\cite{Peruzzo2014} and quantum-approximate-optimization-algorithm (QAOA)\cite{Farhi2014}, spring up and begin to be applied to various financial quantitative tasks\cite{Rebentrost2018, Stamatopoulos2019, Egger2019, Woerner2019, Martin2019, Zoufal2019}. 

Within all sectors of quantitative finance, the Monte-Carlo simulation always plays a significant role\cite{Hull2003, Tuckman2012, Chacko2016}
, as only a few stochastic equations for derivative pricing have found analytical solutions\cite{Black1973, Merton1973}, while most can only be solved numerically by repeating random settings a great many times in an uncertainty distribution ($e.g.$ normal or log-normal distribution), which therefore consumes much time. The quantum amplitude estimation (QAE) algorithm was raised\cite{Brassard2002} in 2002. It is newly suggested as a promising alternative to the Monte Carlo method, as it shows a quadratic speedup comparing to the latter\cite{Rebentrost2018}. So far, applications of QAE for option pricing\cite{Stamatopoulos2019} and credit risk analysis\cite{Egger2019} have been demonstrated. 

Considering the wide use of Monte Carlo simulation and the large variety of pricing models, the involvement of quantum techniques in finance is still at its infancy. Credit derivatives are frequently mentioned financial instruments because of the strong demand for tackling default risks in finance industry. Collateralized debt obligation (CDO) is a multi-name credit derivative backed on a pool of portfolios of defaultable assets (loans, bonds, credits etc.). CDO then packages the portfolio into several tranches with different returns and priorities to suffer the default loss\cite{Chacko2016}. CDO can effectively protect the senior tranche from the loss, but too many default events in the pool would still make the CDO collapsed, which was the case during the subprime financial crisis in 2008. Many voices were then made for improving the CDO pricing model and strengthening regulations in various aspects. Nonetheless, the CDO itself is a useful credit instrument that can work out and redistribute  credit risks in a very quantitative way, and it is still widely studied in quantitative finance. So far, however, the implementation of complex credit instruments like CDO in quantum algorithms has never been reported.  

\begin{figure}[hbt!]
\includegraphics[width=0.48\textwidth]{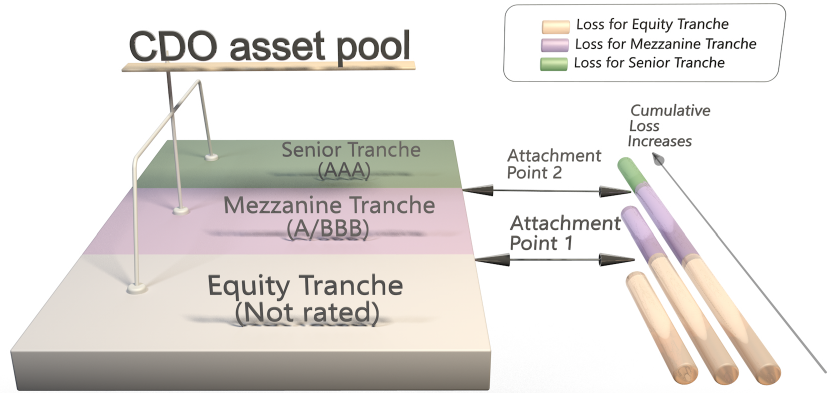}
\caption{\textbf{The CDO tranche structure.} The CDO comprises Equity Tranche (consisting of unrated or lowly rated securities), Mezzanine Tranche (consisting of intermediately rated securities) and Senior Tranche (consisting of highly rated securities), and the tranches have a sequence to bear the loss.   
}\label{fig:CDOTrancheStructure}
\end{figure}

In this work, we present the first quantum circuit implementation for CDO pricing using IBM Qiskit. To address the correlations among a large number of assets in the CDO pool, we use both the common Gaussian copula model\cite{Li2000} and an improved model, the Normal Inverse Gaussian copula model\cite{BarndorffNielsen1978, BarndorffNielsen1997} that can interpret the skewness and kurtosis of the real markets which the Gaussian distribution cannot portray\cite{Guegan2005, Kalemanova2007, Schlosser2011}. We follow a conditional independence approach to load the correlated distributions in the quantum circuits, and then use quantum comparators and QAE algorithm to calculate the losses in different tranches. We demonstrate the quantum computation results for a CDO that matches the classical Monte Carlo method, suggesting a promising approach for pricing various derivatives.

\section{The CDO structure and pricing models}
\subsection{The CDO tranche structures}
The CDO pool is normally divided into three tranches: the Equity, Mezzanine and Senior Tranche. As shown in Fig.1, when defaults occur, the Equity Tranche investors bear the loss first, then the Mezzanine Tranche investors if the loss is greater than the first attachment point. Only when the loss is greater than the second attachment point, will the Senior Tranche investors lose money. Therefore, Senior Tranche has the priority of receiving principle and interest payment, and the best protection from risk while having the lowest return.  

Let $K_{L_k}$ and $K_{U_k}$ denote the lower and upper attachment point for Tranche $k$, respectively. When defaults occur, the buyer of the Tranche $k$ will bear the loss in excess of $K_{L_k}$, and up to $K_{U_k}-K_{L_k}$. Let $L$ denote the total loss for the portfolio and $L_k$ denote the loss suffered by the holders of Tranche $k$. There is: $L_k = min[K_{U_k}-K_{L_k}, max(0,L-K_{L_k})]$. As there are various default scenarios under some uncertainty distribution, we evaluate the expectation value of the tranche loss $\mathbb{E}[L_k]$ for each Tranche $k$: $ \mathbb{E}[L_k] = \mathbb{E}[min[K_{U_k}-K_{L_k}, max(0,L-K_{L_k})] ] $. Then we can get the fair spread for this tranche denoted as $r_k$: 
\begin{equation}
    r_k=\frac{\mathbb{E}[L_k]}{N_k}=\frac{\mathbb{E}[L_k]}{K_{U_k}-K_{L_k}}   
\end{equation}
where $N_k$ is the notional value of Tranche $k$ of the portfolio, which can be calculated by $K_{U_k}-K_{L_k}$. To arrive at a fair price of a CDO, the return for investors of each tranche should be consistent with the expected loss the investors would bear. Therefore, such a fair spread is considered as the return for this tranche. 

\subsection{The conditional independence approach}
Usually the pool in CDO is a portfolio of correlated assets. Their default events are not independent, which can be modeled using the single-factor Gaussian copula. 

Meanwhile, through years' practice on the Gaussian model, it is found not to well portray the phenomena in real CDO markets, $e.g.$, the `correlation smile'\cite{Guegan2005}. In 2005, the Normal Inverse Gaussian (NIG) model was introduced to CDO pricing. In fact, price volatilities in derivative markets seldom show perfect Gaussian distribution. NIG can flexibly introduce a target skewness and kurtosis which the Gaussian model cannot achieve\cite{Guegan2005, Kalemanova2007, Schlosser2011}. Explanation for NIG distribution and its probability density function (pdf) can be seen in Appendix I. 

For either the Gaussian copula or NIG copula model, both of them can use the conditional independence approach\cite{Rutkowski2014} originally developed by Va\v{s}\'i\v{c}ek \cite{Vasicek1987,Vasicek2002} for the multivariate distribution problems. Consider a portfolio that comprises $n$ assets, each with an independent default risk $X_i$, and a correlation $\gamma_i$ with the systematic risk $Z$. The latent variables $W_i$ can be used: $W_i=\gamma_i Z+\sqrt{(1-\gamma_i^2)}X_i$, where $\gamma_i$s are correlation parameters that can be obtained by calibrating the market data. $W_i$, $X_i$ and $Z_i$ generally follow the same type of undertainty distribution, $i.e.$, the three all follow a Gaussian-type distribution in the Gaussian copula model.

Let $p_i^0$ be the original default probability for asset $i$ that is uncorrelated to $Z$. Via detailed derivation\cite{Rutkowski2014} , the default probability under the influence of $Z$ follows:
\begin{equation}
 p_i(z)=F(\frac{F^{-1}(p_i^0)-\sqrt{\gamma_i}z}{\sqrt{1-\gamma_i}}) 
\end{equation}
 
This Eq.(2) is derived for very general scenarios\cite{Rutkowski2014}. $F$ stands for the distribution function of $Z$, which can be any continuous and strictly increasing distribution function, and in this content, they are Gaussian for the Gaussian copula model or NIG for the NIG copula model. $F^{-1}$ stands for the inverse of distribution $F$.  

Using this conditional independence model, given $p_i(z)$ and $\lambda_i$, the loss that would incur for asset $i$ when default happens, the expected total loss would be: 
\begin{equation}
    \mathbb{E}[L] = \int_{-\infty} ^{\infty} \sum_{i=1}^n \lambda_i p_i(z)f(z){\rm d}z  
\end{equation}
where $f(z)$ is the PDF function of $Z$. For Gaussian distribution with a variance $\sigma$, integrating $Z$ from $-3\sigma$ to $3\sigma$ would cover 99.73\% of the distribution. After obtained the expected total loss from Equation (3), we can refer to Equation (1) to get the tranche loss and hence fulfill this task of CDO pricing for each tranche. More derivations for the conditional independence approach and the Monte Carlo method for calculating the tranche loss are given in Appendix II.

\begin{figure}[t!]
\includegraphics[width=0.48\textwidth]{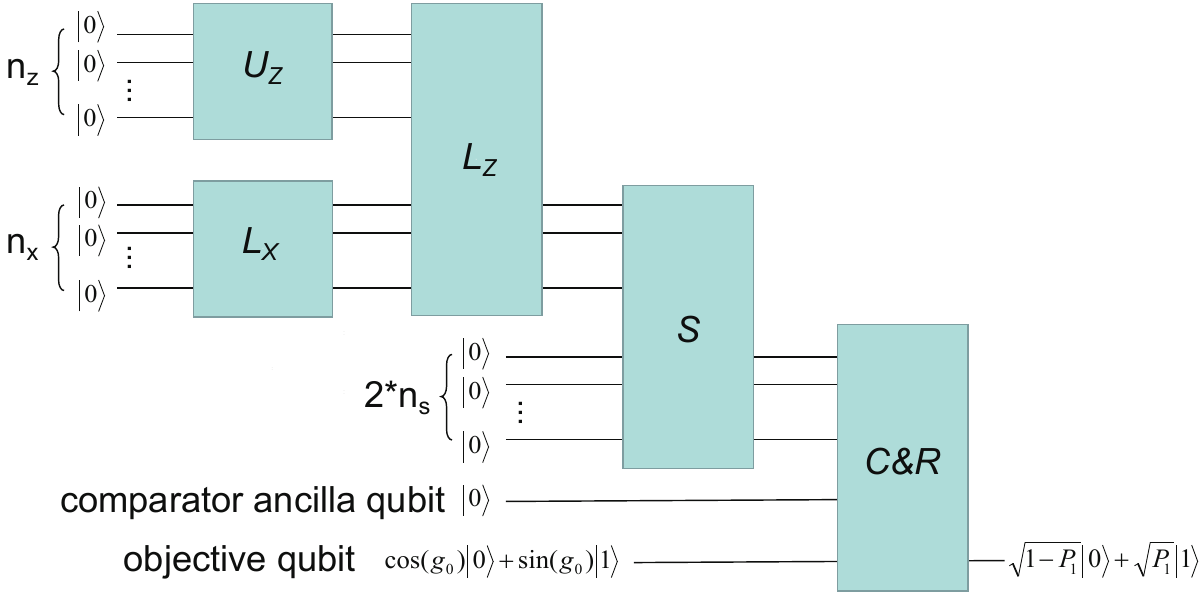}
\caption{\textbf{The quantum circuit framework.} The quantum circuit firstly uses operator $\mathcal{L_X}$ to load the assets with non-correlated independent default risks, uses operator $U_Z$ to load $Z$ distribution and uses operator $\mathcal{L_Z}$ to address the correlation among asset default risks. The total portfolio loss is summed up using operator $\mathcal{S}$. Then it comes to the comparator operator $\mathcal{C}$ and the piecewise linear rotation operator $\mathcal{R}$ to calculate the tranche loss, which is related to $P_1$, the probability of the objective qubit at the state $\ket 1$ after rotation. Detailed circuit for each operator is provided in the appendix.}
\label{fig:QuantumCircuitFramework}
\end{figure}

\section{Quantum circuit construction}

The quantum circuit framework is demonstrated in Fig.2. To apply quantum computation for CDO pricing, the primary task is to load the correlated default risk for each asset of the portfolio into the quantum circuit. Either the Gaussian or NIG model can be loaded following a previous circuit approach\cite{Egger2019} for the conditional independence model. This involves the operator $\mathcal{L_X}$, $\mathcal{U_Z}$ and $\mathcal{L_Z}$, and then sum up the total loss using operator $\mathcal{S}$. 

\subsection{Load uncorrelated default using operator $\mathcal{L_X}$}
We firstly load the uncorrelated asset default event $X_i (i=1,2,...,n_x)$ using linear $Y$-rotation gate. The default probability $p_i$ for each asset $i$ can be obtained from its historical performance. The operator $\mathcal{L_X}$ involves $n_x$ qubits to load the $n_x$ independent assets. For each of the $n_x$ qubits, operator $\mathcal{L_X}$ inputs the initial state $\ket 0^{\otimes n_x}$ and outputs the state:
\begin{equation}
\ket \Psi_{\mathcal{L_X}}=(\sqrt{1-p_i^0}\ket 0+\sqrt{p_i^0}\ket 1)^{\otimes n_x}
\end{equation}
so that the probability for state $\ket 1$ encodes the default probability $p_i^0$. See Appendix III for the circuit for operator $\mathcal{L_X}$.

\subsection{Load $Z$ distribution using operator $\mathcal{U_Z}$}
We need to note that the Gaussian or NIG distribution for systematic risk $Z$ has to be loaded using operator $\mathcal{U_Z}$ before operator $\mathcal{L_Z}$. We use $n_z$ qubits to discretize the distribution to $2^{n_z}$ slots. The $y$ axis for these slots is the probability of $Z$, $i.e.$, the PDF function $f(z)$. For Gaussian distribution function, the $f(z)$ values can be loaded using the built-in codes of uncertainty model in Qiskit, and we contribute the similar codes for NIG distribution. Essentially, operator $\mathcal{U_Z}$ inputs the state $\ket 0^{\otimes n_z}$ for these ${n_z}$ qubits, and outputs $\ket \Psi_{\mathcal{U_Z}}$, a superposition of $2^{n_z}$ entangled states:
\begin{equation}
\ket \Psi_{\mathcal{U_Z}}=\sum_{z=0}^{2^{n_z}-1} \sqrt{f(z)}\ket z 
\end{equation}
where $z$ is an integer in the binary form, e.g. for $n_z$=3, $\ket z$ ranges from $\ket {000}$ to $\ket {111}$, corresponding to -3$\sigma$ to 3$\sigma$ of the Gaussian distribution. Operator $\mathcal{U_Z}$ is constructed via a series of controlled-not gates and unitary rotations. See Appendix IV for the circuit of operator $\mathcal{U_Z}$ and a brief derivation via matrix calculations.

\subsection{Load correlated default using operator $\mathcal{L_Z}$}
Then we need to load the default correlation using operator $\mathcal{L_Z}$. In short, operator $\mathcal{L_Z}$ inputs $\sum_{z=0}^{2^{n_z}-1} \sqrt{f(z)}\ket z$ and $(\sqrt{1-p_i^0}\ket 0+\sqrt{p_i^0}\ket 1)^{\otimes n_x} $ from $\mathcal{U_Z}$ and $\mathcal{L_X}$, respectively, and outputs the further entangled state: 
\begin{equation}
\ket \Psi_{\mathcal{L_Z}}=\sum_{z=0}^{2^{n_z}-1}\sqrt{f(z)}\ket z(\sqrt{1-p_i(z)}\ket 0+\sqrt{p_i(z)}\ket 1)^{\otimes n_x}
\end{equation} 
We use affine mapping\cite{Egger2019} to encode the influence of $Z$ value for the lower $n_x$ qubits. For instance, with $n_z$=3 qubits, for $Z=4=1*2^0+0*2^1+1*2^2-1$, Qubit 1 and Qubit 3 turn on their controlled gates, while Qubit 2 does not switch on its controlled gate, so that the value $Z=4$ is considered for the $n_x$ qubits, and there a probability of $f(4)$ for $Z$ being 4. Meanwhile, there are also many linear $Y$-rotation gate $R_Y(z)$ working on the $n_x$ qubits, which changes the probability for state $\ket 1$ from $p_i^0$ to $p_i(z)$. The expression for $p_i(z)$ as a function of $z$ and the correlation-free $p_i$ just follows Eq.(2), which derives the slope and offset for the rotation gate for operator $\mathcal{L_Z}$, $i.e.$ ${\rm sin}^{-1}(\sqrt{p_i(z)})=\emph{slope}*z+\emph{offset}$. See derivation of slope and offset in Appendix V. The quantum circuit for operator $\mathcal{L_Z}$ is provided in Appendix VI. 

\subsection{Load total loss using operator $\mathcal{S}$}
Furthermore, we set an operator $\mathcal{S}$ to sum up the loss due to all default events in this asset pool. The sum of loss equals to $\sum a_i\lambda_i$, where $\lambda_i$ is the loss given default for asset $i$, and $a_i$ is 1 if asset $i$ default and is 0 vise versa. The probability for $a_i=1$ is just $p_i(z)$ given by operator $\mathcal{L_Z}$. The maximum loss would be $\sum \lambda_i$ when all assets default, so ensuring the maximum loss to be encoded needs $n_s$ qubits that $\sum \lambda_i\leq 2^{n_s}-1$. The operator $\mathcal{S}$ uses $2n_s-1$ qubits following the previous design of sum operator\cite{Egger2019}. It inputs the state: $\ket \Psi_{\mathcal{L_Z}}\ket 0^{\otimes 2n_s-1}$, and outputs the state: 
\begin{equation}
\ket \Psi_{\mathcal{S}}=\ket \Psi_{\mathcal{L_Z}}\ket {\sum_{i=1}^{n_x} p_i(z)\lambda_i}\ket c^{\otimes n_s-1}
\end{equation}
The first $n_s$ qubits are used to load the sum of loss $\sum_{i=1}^{n_x} p_i(z)\lambda_i$. The next $n_s-1$ qubits are used as the carry qubits $\ket c$. 
See details on the circuit in Appendix VII.  

In short, the output after operator $S$ is consistent with the expression for total loss given in Eq.(3). The next step is to compare the total loss with the attachment points for each tranche and work out the tranche loss. 

\begin{figure*}[hbt!]
\includegraphics[width=0.85\textwidth]{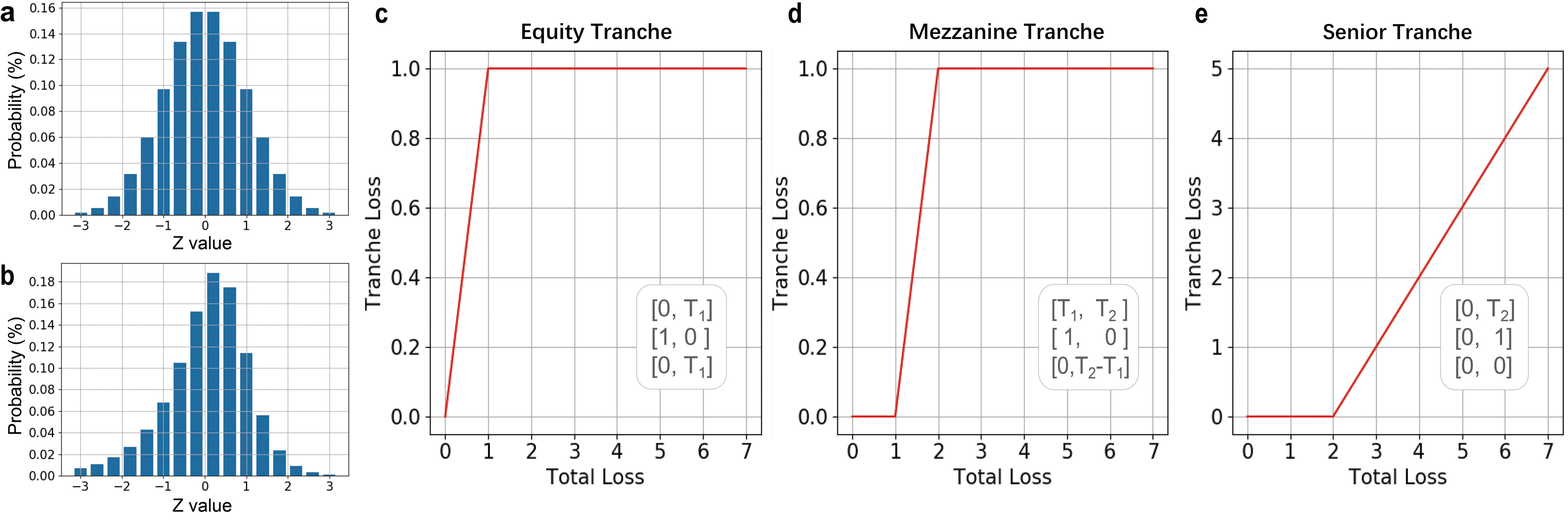}
\caption{\textbf{Systematic risk distribution and the tranche loss functions} (\textbf{a-b}) The probabilities of $2^{n_z}$ different $z$ values using $n_z$ qubits follow the Gaussian distribution(mean=0, variance=1) in (\textbf{a}) and the NIG distribution (skewness=1, kurtosis=6, mean=0, variance=1) in (\textbf{b}). For both distributions, the range is from -3*variance to 3*variance. (\textbf{c-e}) The tranche loss as a function of cumulative loss for (\textbf{c}) Equity Tranche, (\textbf{d}) Mezzanine Tranche, and (\textbf{e}) Senior Tranche. In the white box in (\textbf{c-e}), the first, second and third array respectively show the breakpoints, slopes and offsets for this tranche. $T_1$ is the attachment point between Equity and Mezzanine Tranche, while $T_2$ is that between Mezzanine and Senior Tranche. }
\label{fig:CDOTrancheStructure}
\end{figure*}

\subsection{Load tranche loss using operator $\mathcal{C\&R}$ }
We use the comparator operator $\mathcal{C}_{L_k}$ ($k$=1, 2 and 3) to compare the sum of loss with the fixed lower attachment point $K_{L_k}$ for each Tranche $k$. The comparator has been used to compare the underlying asset value with the striking price for option pricing in a recent work\cite{Stamatopoulos2019}.The operator $\mathcal{C}_{L_k}$ would flip the comparator ancilla qubit from $\ket 0$ to $\ket 1$ if $L(z)$, the sum of loss under the systematic risk $Z$, is higher than $K_{L_k}$, and would keep $\ket 0$ otherwise.

Meanwhile, a piecewise linear rotation operator $\mathcal{R}$ will also rotate the state of an objective qubit under the control of the comparator ancilla qubit. The operator $\mathcal{C\&R}$ inputs the state $\ket \Psi_{\mathcal{S}}\ket 0 ({\rm cos}(g_0)\ket 0+{\rm sin}(g_0) \ket 1)$ and outputs the state $\ket \Psi_{\mathcal{C\&R}}$: 
\begin{equation}
\begin{cases}
\ket \Psi_{\mathcal{S}}\ket 0 ({\rm cos}(g_0)\ket 0+{\rm sin}(g_0) \ket 1) & \text{if}~L(z)\leq K_L\\
\ket \Psi_{\mathcal{S}}\ket 1 ({\rm cos}(g_r)\ket 0+{\rm sin}(g_r) \ket 1) & \text{if}~L(z)>K_L\\
\end{cases}
\end{equation}
where $g0=\frac{\pi}{4}-c$, and $c$ is a scaling factor.  $g_r=g_0+g_z$, where $g_z$ can be implemented using controlled Y-rotations, and it is mapped to integer value $Z \in \{0, \ldots , 2^{n_z}-1\}$. Note that there is an upperbound breakpoint $K_U$ as well, so we set another comparator operator $\mathcal{C}_U$ that encodes $K_U$, and $g_z$ finally reads as:
\begin{equation}
g_z= 2c\frac{min(L(z), K_U)-K_L}{K_U-K_L}
\end{equation} 
With such settings $g_z$ would be in the range $\{0, 2c\}$, and by choosing a small scaling parameter $c$, which is generally set as 0.1 in this work, we can ensure ${\rm sin}(g_0+g_c)$ in a monotonously increasing regime. See Appendix VIII for the quantum circuit of operator $\mathcal{C\&R}$.

Then the probability at state $\ket 1$ is expressed as $P_1$, and it is found to have a relationship with the tranche loss:
\begin{equation}
P_1=(\frac{1}{2}-c)+\frac{2c}{K_{U_k}-K_{L_k}}(\mathbb{E}[L_k])
\end{equation}
where $\mathbb{E}[L_k]$ is the expectation of loss for a certain tranche $k$, for instance, the loss for Equity Tranche when setting $K_{L_1}$ and $K_{U_1}$. See detailed derivation for Eq.(6) in Appendix IX.

\subsection{Calculate tranche loss using QAE}
Then it comes to the issue how to read the value of $P_1$. Quantum Amplitude Estimation (QAE) has been demonstrated as a good alternative to Monte Carlo simulation \cite{Montanaro2017} for finance pricing\cite{Rebentrost2018, Stamatopoulos2019, Woerner2019, Egger2019}. In this work, QAE that estimates $P_1$ allows us to obtain the CDO tranche loss and return. The canonical QAE algorithm was raised in 2002 \cite{Brassard2002}, which is to map the amplitude to be estimated ($P_1$ in this case) to the discretized value using $m$ additional qubits via controlled rotations and inverse Quantum Fourier Transform (QFT). QAE can achieve quadratic speedup, but involvement of inverse QFT requires exponentially increasing circuit depths. Therefore, there arising a series of adapted QAE methods to reduce the complexities of quantum circuits\cite{Grinko2019,Suzuki2019, Aaronson2019}. Here we implement an interative QAE\cite{Grinko2019} (IQAE) for our tranche pricing task. IQAE was raised in late 2019 and it has now become widely used, $e.g.$, the Qiskit module has replaced the canonical QAE with QAE for many tutorial modules. The methods for both canonical QAE and iterative QAE are provided in Appendix X.  
 
\section{Result analysis}

We consider an example to show the pricing for CDO tranches. As listed in Table I, the CDO pool has four assets, each showing a default probability $p^0_{i}$, a sensitivity to the systematic risk $\gamma_i$ and a loss given default $\lambda_{i}$. 

The CDO is divided into three tranches: the Equity, Mezzanine and Senior Tranches. Values for the lower attachment point $K_{L_k}$ and upper attachment point $K_{U_k}$ for three tranches are provided in Table II. 

For this task, we need $n_x=4$ qubits to represent the four assets in operator $\mathcal{L_X}$, and $n_z=4$ qubits in operator $\mathcal{L_Z}$ to make $2^4=16$ slots for the uncertainty distribution of systematic risk $Z$. We implement Gaussian (Fig.3a) and NIG (Fig.3b) distribution for $Z$. 

For NIG distribution, by setting the parameters given in Appendix I, it shows a skewness of 1 and kurtosis of 6, which are consistent with a real CDO market\cite{Guegan2005}. Comparing with Gaussian distribution, this is narrower and centered to the left. 

The step after loading distribution is to calculate the cumulative loss. The maximal loss is $\sum \lambda_i=7$ for this portfolio. Therefore, we can use $n_s=3$ qubits to encode the total loss in the weighted sum operator $\mathcal{S}$.  

The pricing of the tranche loss is similar to the call option pricing, where there is a linear `payoff function' that goes up from zero after the option striking price or, for the CDO tranche, the attachment point. The tranche loss as a function of the total cumulative loss is given in Fig.3c-e for this specific example. See Appendix XI for how to set the input parameters ($e.g.$, `breakpoint') for the piecewise linear rotation function.

\begin{table}[H]
\caption{\textbf{The relevant parameters for each asset.} }
\begin{center}
\begin{tabular}{p{1.2cm}p{1cm}p{1cm}p{1cm}}
\hline\noalign{\vskip 0.5mm}
		\hline\noalign{\vskip 0.14cm}
Asset $i$ & $\lambda_{i}$ & $p^0_{i}$ & $\gamma_i$ \\
\hline
1 & 2 & 0.3 & 0.05 \\ 
2 & 2 & 0.1 &0.15 \\ 
3 & 1 & 0.2 & 0.1 \\ 
4 & 2 & 0.1 & 0.05\\
\hline\noalign{\vskip 0.5mm}
		\hline\noalign{\vskip 0.14cm}		
\end{tabular}
\end{center}
\caption{\textbf{The attachment points for each tranche.}}
\begin{center}
\begin{tabular}{ p{2.5cm}p{2cm}p{2cm} }
\hline\noalign{\vskip 0.5mm}
		\hline\noalign{\vskip 0.14cm}
Tranche Name & Lower $K_{L_k}$ & Upper $K_{U_k}$ \\
\hline
Equity & 0 & 1 \\ 
Mezzanine & 1 & 2 \\ 
Senior & 2 & 7 \\ 
\hline\noalign{\vskip 0.5mm}
		\hline\noalign{\vskip 0.14cm}		
\end{tabular}
\end{center}
\end{table}

We then use IQAE to estimate $P_1$ and convert it to the tranche loss according to Eq.(4). We use the QASM cloud backend that is in the Noisy Intermediate-Scale Quantum (NISQ) environment. Fig.4 demonstrates the calculated tranche loss for an NIG distribution (Fig.3b) using IQAE with $\epsilon=0.001$ and $\alpha=0.05$, the expected wavefunction results from the quantum circuit and the classical Monte Carlo method. The results for different approaches match well. When $Z$ follows Gaussian distribution (Fig.3a), consistent results have also been obtained, as shown in Fig.A11 in the appendix. Still, the NIG results slight differs from the Gaussian results with a relatively lower tranche loss, especially for the senior tranche loss, which is 0.2233 for NIG and 0.2301 for Gaussian distribution, both obtained via the matrix calculation result for related quantum circuits. This can be due to the skewed distribution for NIG, which makes more positive $Z$ values than the Gaussian one, so that expected total loss will be relatively lower considering a negative $p_i(z)-z$ relationship given in Eq.(2) and a positive $L(z)-p_i(z)$ relationship given in Eq.(4). Therefore, if the real market follows an NIG distribution while we use Gaussian distribution to model it, we would over estimate the expected tranche loss.

With the calculated tranche loss, we can price the CDO tranche return according to Eq.(1). The notional value $N$ for the Equity Tranche, Mezzanine Tranche and Senior Tranche is 1, 1, and 5, respectively, by calculating $K_U-K_L$ for each tranche. For Equity, Mezzanine and Senior Tranche, the expected tranche loss via IQAE gives 52.0\%, 41.7\% and 23.8\%. Then the tranche return for these tranches are 52.0\%, 41.7\% and 4.76\%, respectively. The low return for the Senior Tranche is consistent with the practice in reality. Such a low value is firstly due to the last sequence to bear the loss, and secondly owing to its large notional value, which is normally above 80\% of the sum for the three tranches. See more discussion on tranche return in practice in Appendix XII. 

\begin{figure*}[hbt!]
\includegraphics[width=0.9\textwidth]{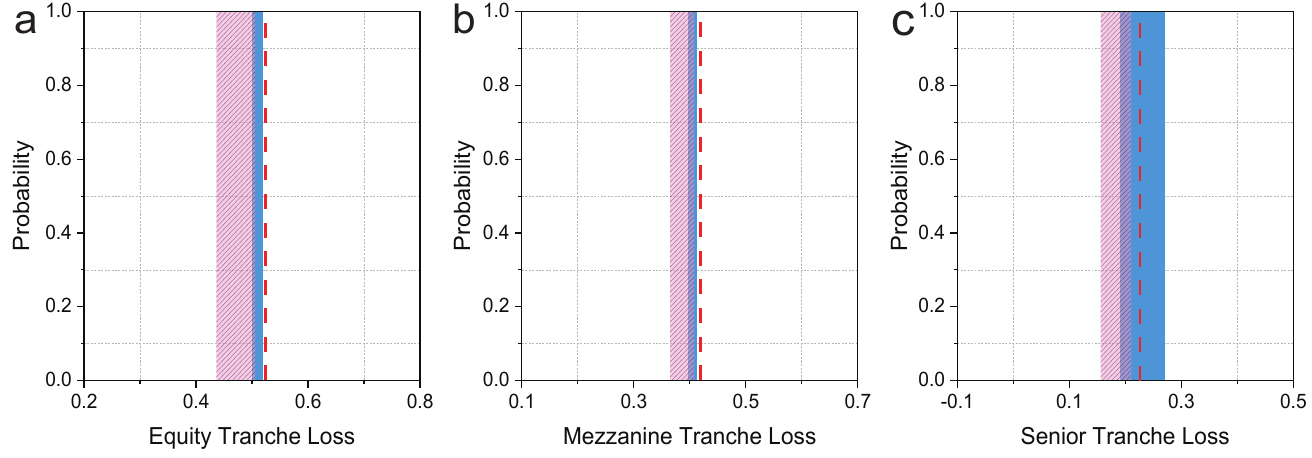}
\caption{\textbf{CDO tranche loss with $Z$ under the NIG distribution.} The calculated loss for (\textbf{a}) Equity Tranche, (\textbf{b}) Mezzanine Tranche, and (\textbf{c}) Senior Tranche. $Z$ follows the NIG distribution depicted in Fig.3b. Blue bars indicate quantum computation results with $\epsilon=0.002$ and $\alpha=0.05$ for IQAE using QASM simulator. Red dashed lines indicate definite matrix calculation result of the quantum circuit. The pink shading areas indicate the Monte Carlo results, which are obtained by finding the range of 20 sets of Monte-Carlo simulations, each set having 1000 random repetitions.}
\label{fig:CDOTrancheLoss}
\end{figure*}

We further conduct a robustness analysis on the quantum computation method for CDO tranche pricing, with details shown in Appendix XIII. We have noticed that the scaling factor $c$ is introduced in the operator $\mathcal{C\&R}$, and it has to be small enough to make the approximation ${\rm sin^2}(g_0)=\frac{1}{2}-c$ satisfied. Therefore, a smaller value of $c$ tends to be more accurate.  The parameters for iterative QAE are investigated as well, including the confidence interval parameter alpha and the precision parameter epsilon. We find that $\alpha$ does not have a prominent influence on the range of given result, while $\epsilon$ severely impacts the range, and the result is satisfactory only when $\epsilon$ goes down to 0.002 or below. For all these quantum computing parameters, the senior tranche is most sensitive to these changes. This can be caused by its last sequence to bear loss, while little fluctuation would not change the result for other tranches, a slight decrease of total loss can possibly exempt the loss due for the senior tranche. We also introduce up to $50\%$ fluctuations of either the independent probability of default $p_i^0$ or the correlation to systematic risk, $\gamma_i$, and find that the fluction of $p_i^0$ has a stronger influence on the tranche loss result comparing to $\gamma_i$. We show that each tranche loss shows an increasing and decreasing trend with $p_i^0$ and $\gamma_i$, respectively, which is consistent with the theoretical conditional independence appoach in Eq.(2). In addition, an analysis is given in Appendix XIV to show how different operators scale with $n_x$, $n_z$ and $n_s$. It suggests that $\mathcal{U_Z}$ and $\mathcal{S}$ consumes heavy circuit depth and it is worth investigation for further optimization for those operators.

\section{Discussion and Conclusion}

The CDO is a relatively advanced and complex structured finance product, and the credit market plays a significant role in the finance industry. Therefore, despite there were some disputes on CDOs during the 2008 financial crisis, CDOs are still widely studied products in quantitative finance, and are being improved with various financial models. In this work, we implement the normal inverse Gaussian model that is now regarded as advantageous over the Gaussian model. There is also the variance gamma model that was first applied to option pricing\cite{Madan1998} and later found to be a good model for CDO pricing\cite{Moosbrucker2006}. Such improved models can also be calculated via quantum computation. 

Note that the quantum adaption of generative adversarial network\cite{Goodfellow2014,Lloyd2018} has now been considered as an effective way to load any distribution in quantum circuits\cite{Zoufal2019} and can be applied to more finance models. Besides, the parameter shift rule\cite{Li2017, Schuld2018} has been raised to solve the issue of encoding gradients in quantum circuits, which facilitates the mapping of machine learning techniques in quantum algorithms. Furthermore, the trendy variational quantum algorithms that are suitable for NISQ environment, and the alternative approach using quantum annealing\cite{Cohen2020}, may work on a large variety of optimization tasks in finance. In all, there's much room to explore for quantum computation in finance applications. 
\bigskip
\begin{acknowledgments}
H.T. thanks Prof. Stephen Schaefer's previous help for studies on fixed income and interest rate derivative at London Business School. The authors thank Jian-Wei Pan for helpful discussions. This research was supported by National Key R\&D Program of China (2019YFA0308700, 2017YFA0303700), National Natural Science Foundation of China (61734005, 11761141014, 11690033, 11904229), Science and Technology Commission of Shanghai Municipality (STCSM) (17JC1400403), and Shanghai Municipal Education Commission (SMEC) (2017-01-07-00-02- E00049). X.-M.J. acknowledges additional support from a Shanghai talent program.

\textbf{Author Contributions.}
H.T. and X.-M.J. conceived and supervised the project. H.T. and A.P. designed the scheme. A.P. wrote the Qiskit code. H.T. did Monte Carlo simulation. H.T., A.P., L.F.Q., T.Y.W., J.G. and X.M.J. analyzed the data and presented the figures. H.T., A.P. and T.Y.W. enriched the quantum circuit analysis in the appendix. H.T. wrote the paper, including the appendix, with input from all the other authors.
\textbf{Competing Interests.}
The authors declare no competing interests.
\textbf{Data Availability.}
The data that support the plots within this paper and other findings of this study are available from the corresponding author upon reasonable request.
\end{acknowledgments}

\begin{appendix}
\clearpage
\newpage
\twocolumngrid

\section*{Appendix I The model of Normal Inverse Gaussian Distribution}
\setcounter{table}{0}
\setcounter{equation}{0}
\setcounter{figure}{0}

\renewcommand{\thetable}{{A}\arabic{table}}
\renewcommand{\theequation}{{A}\arabic{equation}}
\renewcommand{\thefigure}{{A}\arabic{figure}}

The Normal Inverse Gaussian (NIG) distribution mixes the normal Gaussian distribution and inverse Gaussian distributions[Ref A1].
 
The word `inverse' in the name needs to be explained. While normal distribution reflects the location distribution under Brownian motion at a certain time, the inverse Gaussian distribution shows the time distribution when the Brownian motion moves to a certain location, so inverse suggests an inverse way in viewing location and time. 

Firstly, for a random variable $Y$ that has inverse Gaussian distribution, its density of function is of the form:
\begin{equation}
f_{IG}(y;\alpha, \beta)= \frac{\alpha}{\sqrt{2\pi\beta}}y^{-3/2}\rm exp(-\frac{(\alpha-\beta y)^2}{2\beta y})
\end{equation}

Then if a random variable $X$ satisfies the following requirement with parameters $\alpha$, $\beta$, $\mu$ and $\delta$, it follows the Normal Inverse Gaussian distribution $\mathcal{NIG}(x; \alpha, \beta, \mu, \delta)$:
\begin{equation}
\begin{split}
X|Y =y \sim \mathcal{N}(\mu+\beta y,y)\\
Y \sim \mathcal{IG}(\gamma, \gamma^2) ~{\rm wi}&{\rm th}~ \gamma=\sqrt{\alpha^2-\beta^2}
\end{split}
\end{equation}

The full expression of the probability density function for NIG distribution is a bit complicated:
\begin{equation}
\begin{split}
\mathcal{NIG}(x; \alpha, \beta, \mu, \delta)=\\a(\alpha, \beta, \mu, \delta)q(&\frac{x-\mu}{\delta})^{-1}K_1(\delta\alpha q(\frac{x-\mu}{\delta}))e^{\beta x}
\end{split}
\end{equation}
where the function $q$ follows: $ q(x)=\sqrt{1+x^2}$, and $a$ is the function with variables $\alpha$, $\beta$, $\mu$ and $\delta$:
\begin{equation}
a(\alpha,\beta, \mu, \delta)=\pi^{-1}\alpha {\rm exp} (\delta\sqrt{\alpha^2-\beta^2-\beta\mu})
\end{equation}
with parameters satisfying: $0\leq |\beta| <\alpha$ and $\delta>0$, and $K_1$ is the first index of the Bessel function of the third kind:
\begin{equation}
K_1(x)=x\int_1^{\infty} {\rm exp}(-xt)\sqrt{t^2-1}dt
\end{equation}

The parameter $\alpha$ is related to steepness, $\beta$ to symmetry, $\mu$ to location and $\delta$ to scale. In order to realize the NIG distribution (mean=0, variance=1, skewness=1, and kurtosis=6), the parameters need to be set as follows[Ref A2]:
$\alpha$ =-1.6771, $\beta$=0.75, $\mu$=-0.6, and $\delta=1.2$.

It's worth noting that despite of the complexity of NIG distribution functions, they can be conveniently implemented using the built-in functions in Scipy. We then write the uncertainty model and conditional independence model for the NIG distribution and contribute it to the Qiskit package, in the same folder with those for the Gaussian distribution ($\backslash$qiskit$\backslash$aqua$\backslash$components$\backslash$uncertainty$\_$models). 

\begin{figure}[b!]
\includegraphics[width=0.45\textwidth]{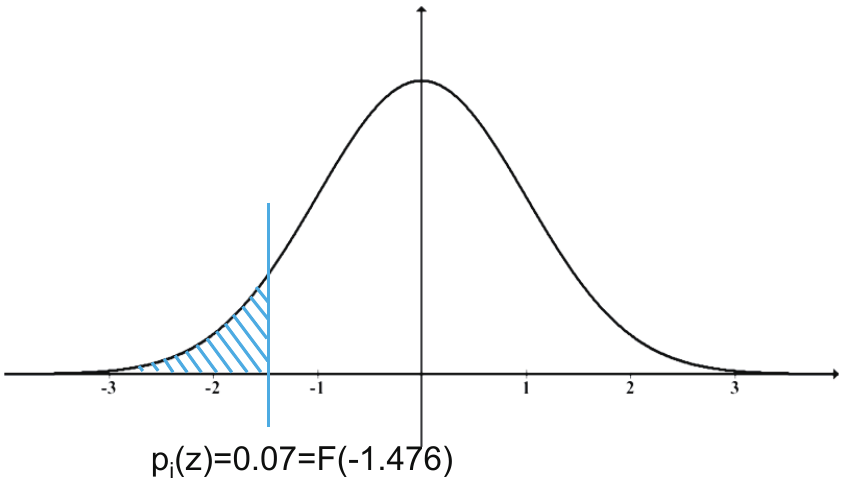}
\caption{\textbf{The criterion of an asset default event.} The blue shading area of the Gaussian distribution shows an area of 0.07, considering the whole curve has formed an area of 1. The area of 0.07 is the value of $p_i(z)$, and it corresponds to $F(-1.476)$. If a random value $q$ that follows a Gaussian distribution for asset $i$ is smaller than -1.476, that is, $F(q)<F(-1.476)$ and the formed shading area is smaller than 0.07, then we regard this asset defaults. It's clear that this asset would default with a probability of $p_i(z)=0.07$. }
\label{fig:assetdefaul}
\end{figure}

~\\
\section*{Appendix II The conditional independence approach and the Monte Carlo method}

Usually the pool in CDO is a portfolio of correlated assets, where default event for each asset is not independent. 
Consider a portfolio that comprises $n$ assets, according to the conditional independence approach, we can analyze the mutually correlated assets by considering each asset has an independent default risk $X_i$ and a correlation $\gamma_i$ with the systematic risk $Z$.  We can use the mutually latent variables $W_i$ to describe this:  
\begin{equation}
W_i=\gamma_i Z+\sqrt{(1-\gamma_i^2)}X_i 
\end{equation}
where $W_i$ $Z$ and $X_i$ follow the $n$-dimensional distribution function $F$ that has marginal function $F_1$, $F_2$, ..., $F_n$. $\gamma_i$ are correlation parameters that can be obtained by calibrating the market data. 
Let ($p_1^0$, $p_2^0$, ..., $p_i^0$, ..., $p_{n_x}^0$) be the unconditional probability of default for each asset. Then asset $i$ is regarded to default when $W_i<F_1^{-1}(p_i^0)$ if there is no correlation among $W_i$, where $F^{-1}$ is the inverse of distribution function $F$. The $n$-copula can describe the joint default profile: 
\begin{equation}
\begin{split}
C(u_1,...u_n)=F(F_1^{-1}(p_1^0),...,F_n^{-1}(p_{n_x}^0))\\=P(W_1<F_1^{-1}(p_1^0),..., W_n&<F_n^{-1}(p_{n_x}^0))
\end{split}
\end{equation}

\begin{table}[t!]
\caption{\textbf{The asset parameters in a simpler case.} }
\begin{center}
\begin{tabular}{p{1.2cm}p{1cm}p{1cm}p{1cm}}
\hline\noalign{\vskip 0.5mm}
		\hline\noalign{\vskip 0.14cm}
Asset $i$ & $\lambda_{i}$ & $p^0_{i}$ & $\gamma_i$ \\
\hline
1 & 1 & 0.1 & 0.1 \\ 
2 & 1 & 0.2 &0.15 \\ 
3 & 1 & 0.3 & 0.05 \\ 
\hline\noalign{\vskip 0.5mm}
		\hline\noalign{\vskip 0.14cm}		
\end{tabular}
\end{center}
\caption{\textbf{The attachment points in a simpler case.}}
\begin{center}
\begin{tabular}{ p{2.5cm}p{2cm}p{2cm} }
\hline\noalign{\vskip 0.5mm}
		\hline\noalign{\vskip 0.14cm}
Tranche Name & Lower $K_{L_k}$ & Upper $K_{U_k}$ \\
\hline
Equity & 0 & 1 \\ 
Mezzanine & 1 & 2 \\ 
Senior & 2 & 3 \\ 
\hline\noalign{\vskip 0.5mm}
		\hline\noalign{\vskip 0.14cm}		
\end{tabular}
\end{center}
\end{table}

As demonstrated in Eq.(A6), in fact, the latent variables $W_i$s are correlated due to a correlation to the systematic risk $Z$. Using $W_i$ as the variability in obligors’ asset values, then there is a pairwise correlation between obligors’ asset values: ${\rm Corr}(W_i, W_j) = \sqrt{\gamma_i\gamma_j}$. We need to consider $p_i(z)$, which is conditional probability of default, $i.e.$, the probability of default of asset $i$ conditional on realisation $z \in \mathcal R$ on systematic risk $Z$, and it can be expressed as:
\begin{equation}
\begin{split}
 p_i(z)=P(W_i<F_1^{-1}(p_i^0)|Z=z)\\
= P(\gamma_i Z+\sqrt{(1-\gamma_i^2)}X_i<&F_1^{-1}(p_i^0)|Z=z)\\
= P(X_i<\frac{F^{-1}(p_i^0)-\sqrt{\gamma_i}z}{\sqrt{1-\gamma_i}}) \\
= F(\frac{F^{-1}(p_i^0)-\sqrt{\gamma_i}z}{\sqrt{1-\gamma_i}})
\end{split}
\end{equation}
This becomes what is shown in Eq.(2) of the main text. We can see that this is derived for very general case. It just requires that $W_i$, $X_i$ and $Z$ follow the continuous and strictly increasing distribution functions, and for simplicity, we regard $W_i$, $X_i$ and $Z$ follow the same type of distribution. In this single-factor Gaussian model, the systematic risk $Z$ follows the normal Gaussian distribution, and in the NIG model, $Z$ follows an NIG distribution.

To solve the tranche loss using the Monte Carlo method, we can repeat the $Z$ value for 1000 times, each time with a random value, and the histogram of the 1000 $Z$ values following a certain distribution, e.g., Gaussian or NIG distribution. For each random $Z$, we calculate $p_i(z)$ for each asset $i$ according to Eq.(2) of the main text.  We assume the asset price fluctuation follows the same distribution with $Z$ and get an accumulative probability $F(q)$ for a random value $q$. By comparing $F(q)$ with $p_i(z)$, if $F(q)
\leq p_i(z)$, $i.e.$, $q\leq F^{-1}(p_i(z))$, then we regard this asset defaults. Clearly, there is overall a probability of $p_i(z)$ for this asset to default. This is illustrated in Fig.A1.
 
For each of the 1000 random settings, we calculate the total loss according to Eq.(3) of the main text, and further obtain the tranche loss for each tranche according to Eq.(1). Then we sum up each tranche loss and divide it by 1000, so we get one estimation of the expected tranche loss for each tranche. We repeat estimations for 20 times, so we get a range of estimations for Monte Carlo results. Initially, we just ran Monte Carlo simulation to get one estimation, and now we suggest it's more reasonable to use a range of results as the Monte Carlo result, as it does not give a definite answer inherently.

~\\
\section*{Appendix III Quantum circuit for the operator $\mathcal{L_X}$}
The quantum circuit for operator $\mathcal{L_X}$ simply involves one Y rotation gate $R_Y$ for each of the $n_x$ qubits. The rotation angle $\theta_i$ for $R_Y(\theta_i)$ satisfies: $\theta_i=2{\rm arcsin}(\sqrt {p_i^0})$, so that the probability of measuring qubit $i$ at state $\ket 1$ would be the independent default probability $p_i$ for asset $i$. In order to show the method of circuit construction, we use a simpler CDO asset pool with three assets (see Table A1), with a $p^0$ of 0.1, 0.2 and 0.3 respectively. Fig.A2 shows the operator $\mathcal{L_X}$ circuit with three qubits for these three assets. 

\begin{figure}[t!]
\includegraphics[width=0.16\textwidth]{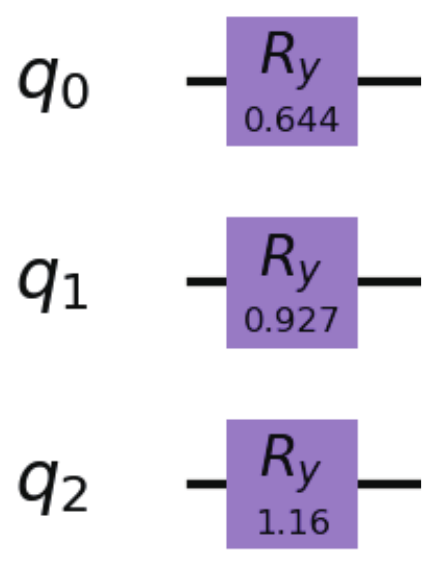}
\caption{\textbf{Quantum circuit for the operator $\mathcal{L_X}$.} It involves an $R_Y$ rotation gate for each of the $n_x=3$ assets.  }
\label{fig:assetdefaul}
\end{figure}

~\\
\section*{Appendix IV Quantum circuit for the operator $\mathcal{U_Z}$}
The quantum circuit for operator $\mathcal{U_Z}$ uses a series of controll-rotation gates and unitary gates to make the state: $\sum_{z=0}^{2^{n_z}-1} \sqrt{f(z)}\ket z$. Fig.A3 shows a simple circuit for $n_z=2$ qubits and plots the probability distribution for $Z$, which roughly follows a Gaussian distribution. The precision of loading distribution will largely improve when increasing $n_z$. 

\begin{figure*}[t!]
\includegraphics[width=0.95\textwidth]{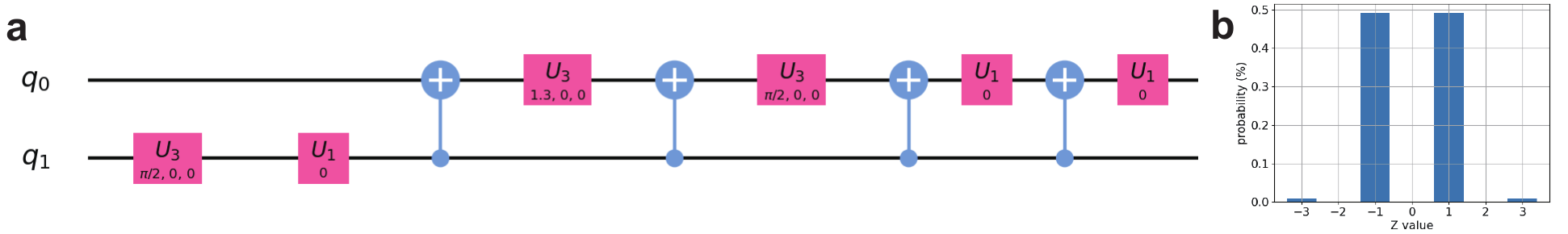}
\caption{\textbf{Quantum circuit and the output state for the operator $\mathcal{U_Z}$.} (\textbf{a}) The quantum circuit of using $n_z=2$ qubits to load the distribution. (\textbf{b}) The output state of operator $\mathcal{U_Z}$. The four bars from left to right correspond to $\ket {q_1q_0}$ of $\ket {00}$, $\ket {01}$, $\ket {10}$ and $\ket {11}$, respectively.}
\label{fig:assetdefaul}
\end{figure*}

\begin{figure*}[t!]
\includegraphics[width=0.99\textwidth]{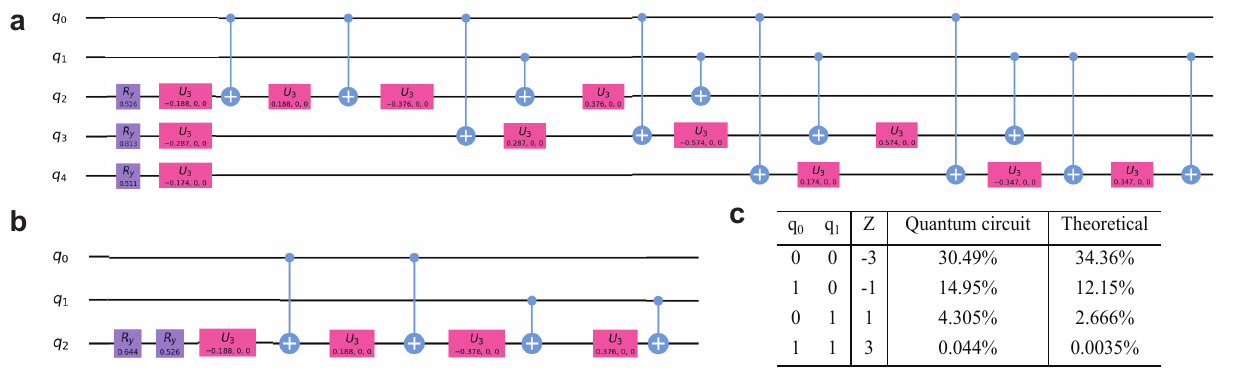}
\caption{\textbf{Quantum circuit for the operator $\mathcal{L_Z}$.} (\textbf{a}) The full circuit for operator $\mathcal{L_Z}$. (\textbf{b}) The circuit for loading conditional default probability $p_i(z)$ for the first asset. (\textbf{c}) $p_i(z)$ for different $Z$ values. `Quantum circuit" refers to the probability for $q_2$ at state $\ket 1$ via matrix calculation for the quantum circuit shown in (\textbf{b}). `Theoretical' refers to the result calculated using Eq.(2) of the main text. }
\label{fig:assetdefaul}
\end{figure*}

~\\
\section*{Appendix V Quantum circuit for the operator $\mathcal{L_Z}$}
The operator $\mathcal{L_Z}$ uses a series of controlled-rotation gate to make the probability at state $\ket 1$ to be $p_i(z)$, the default probability conditional to the systematic risk Z. An example of the quantum circuit for operator $\mathcal{L_Z}$ is given in Fig.A4a, following up the circuit for $\mathcal{L_X}$ and $\mathcal{U_Z}$ show in in Fig. A2 and A3a, respectively, so that this operator $\mathcal{L_Z}$ needs 5 qubits. $\gamma_i$, the correlation to systematic risk $Z$, for the three assets as defined in Appendix III, is set to be 0.1, 0.15 and 0.05, respectively. We can easily see this circuit comprises three similar units, one for each asset. 

We can quickly verify one case of the first asset. As shown in Fig.A4b, the first $R_Y$ gate is actually the $\mathcal{L_X}$ operation that loads initial default probability $p_i^0$, and the following circuit is just a unit of that in Fig.A4a and will output conditional default probability $p_i(z)$. Here we just try discrete $Z$ values one by one, $i.e$, -3, -1, 1 and 1, and do not consider the occuring probability for each $Z$ shown in Fig.A3b. We present the probability at state $\ket 1$, which is just $p_i(z)$, in Fig. A4c for different $Z$ values. The circuit output results are consistent with the expected result from Eq.(2).

\begin{figure*}[t!]
\includegraphics[width=0.9\textwidth]{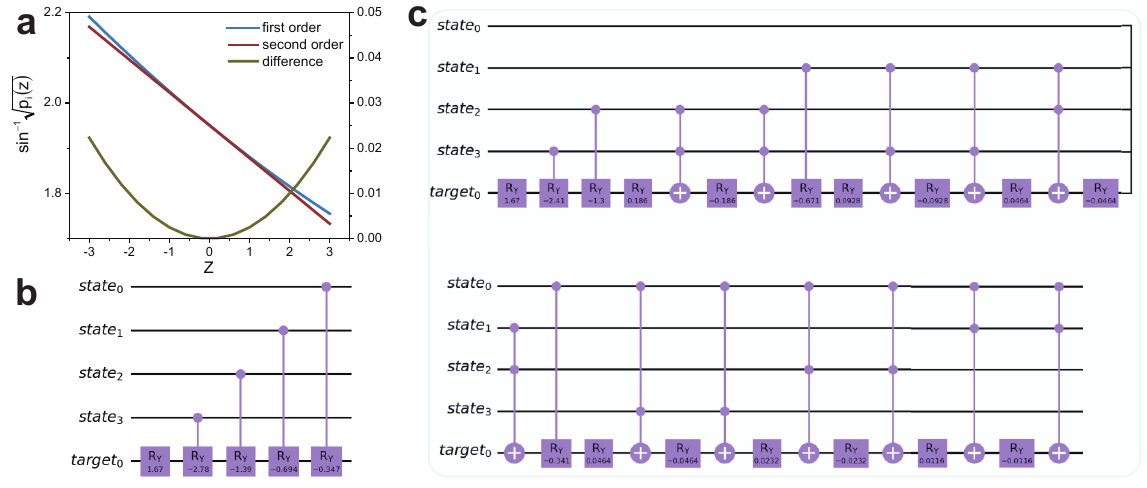}
\caption{\textbf{Second-order Taylor's theorem for ${\rm sin}^{-1}\sqrt{p_i(z)}$.} (\textbf{a}) Comparing the approximation result for ${\rm sin}^{-1}\sqrt{p_i(z)}$ using first-order and second-order Taylor's theorem. (\textbf{b}) The quantum circuit to load the first-order approximation, $y=1.66975-0.3472x$. (\textbf{c}) The quantum circuit to load the second-order approximation, $y=1.66975-0.3472x+0.0058x^2$.}
\label{fig:slopeLz}
\end{figure*}

\begin{figure*}[t!]
\includegraphics[width=0.8\textwidth]{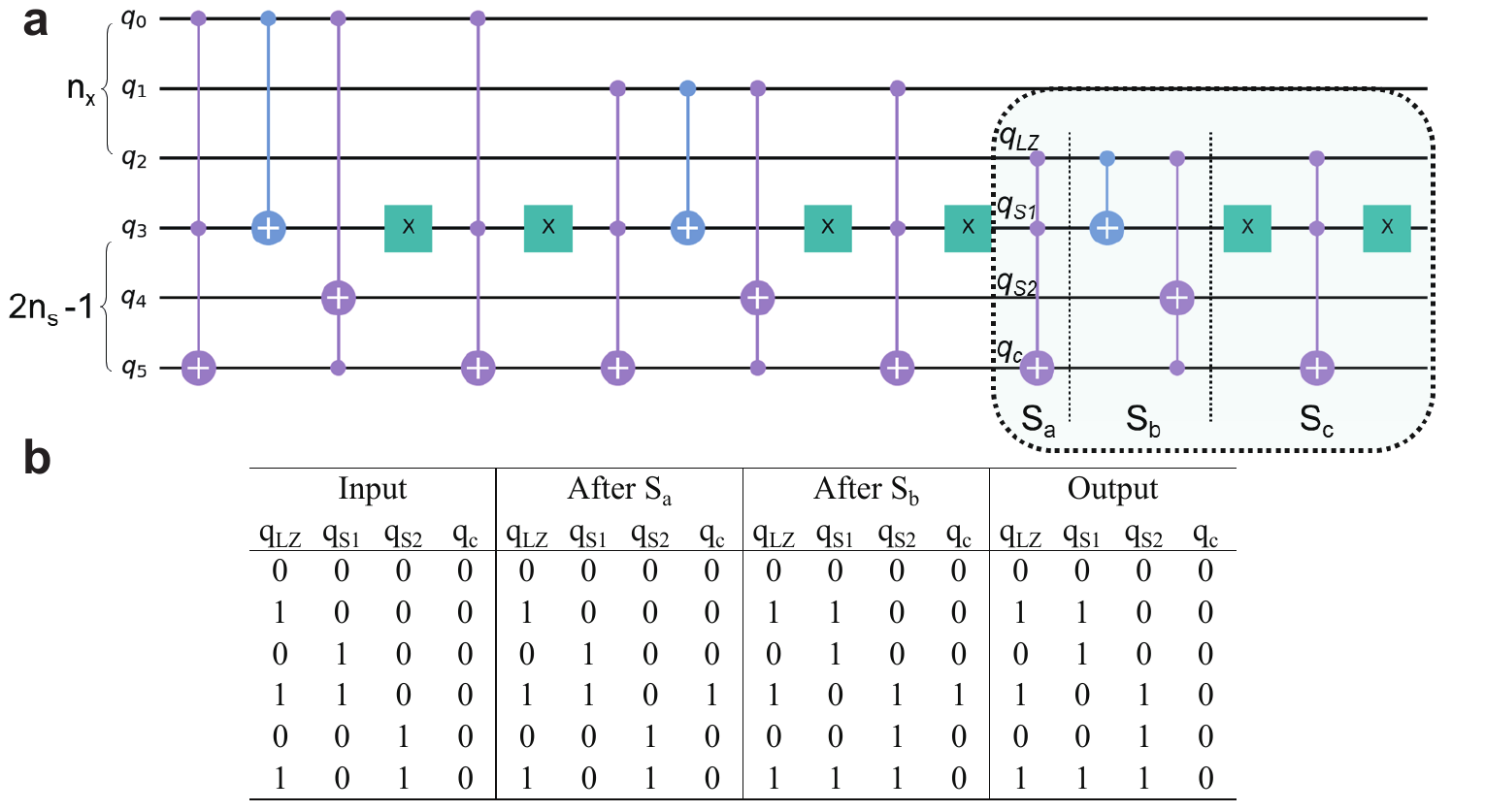}
\caption{\textbf{Quantum circuit for the operator $\mathcal{S}$.}  (\textbf{a}) We consider the loss given default for the three assets as mentioned in Appendix III to V to be $\lambda_1=1$, $\lambda_2=1$ and $\lambda_3=1$ respectively, so that the maximal loss of 3 requires $n_s=2$ qubits for loading the sum ($q_3$ and $q_4$ in this figure) and 1 carry qubit ($q_5$ in this figure). The top three qubits are in fact the $n_x$ qubits that has a probabibity of $p_i(z)$ at $\ket 1$. They are connected to the $2n_s-1$ qubits via controlled rotation gates. The part circled out is just one unit for the sum operator. We denote the four qubits in this unit $q_{LZ}$, $q_{S_1}$, $q_{S_2}$ and $q_c$ respectively. The unit comprises of three components $S_a$, $S_b$ and $S_c$. The truth table for this unit is provided in (\textbf{b}). }
\label{fig:assetdefaul}
\end{figure*}

~\\
\section*{Appendix VI Derive the linear rotation parameters for operator $\mathcal{L_Z}$ and explain affine mapping}

A linear rotation function would use the state qubit $\ket x$ to work on the target qubit $\ket 0$: 
\begin{equation}
\begin{split}
\ket x \ket 0 \rightarrow \ket x ({\rm cos} (\emph{slope}*x+\emph{offset})\ket 0\\ +{\rm sin} (\emph{slope}*x+\emph{offset})\ket 1)
\end{split}
\end{equation}
After the rotation, the probability at the state $\ket 1$ would be the probability that we are interested in, and it is the $z$-tuned default probability $p_i(z)$ in operator $\mathcal{L_Z}$:
\begin{equation}
\sqrt{p_i(z)}={\rm sin} (\emph{slope}*z+\emph{offset})
\end{equation}

so ${\rm sin}^{-1}\sqrt{p_i(z)}$ has to be expressed in the form of $\emph{slope}*z+\emph{offset}$ to get the \emph{slope} and \emph{offset} for the linear rotation quantum gate in operator $\mathcal{L_Z}$.

Combining the expression for $p_i(z)$ in Eq.(2) using the conditional independence model, we have:
\begin{equation}
{\rm sin}^{-1}\sqrt{p_i(z)}={\rm sin}^{-1}\sqrt{F(\frac{F^{-1}(p_i^0)-\sqrt{\gamma_i}z}{\sqrt{1-\gamma_i}})}
\end{equation}
where $F$ is the cumulative distribution function (CDF), and we denote
$\psi =\frac{F^{-1}(p_i^0)}{\sqrt{1-\gamma_i}}$, so the above equation can be simply expressed as: 
\begin{equation}
{\rm sin}^{-1}\sqrt{p_i(z)}={\rm sin}^{-1}\sqrt{F(\psi-\frac{\sqrt{\gamma_i}z}{\sqrt{1-\gamma_i}})}
\end{equation}

The first order Taylor's theorem can be expressed as:
\begin{equation}
g(x)=g(a)+g'(a)(x-a)
\end{equation}
where $g'(a)$ is the first derivative of the function $g(x)$. 

Now let $x$ be $\psi-\frac{\sqrt{\gamma_i}z}{\sqrt{1-\gamma_i}}$, and $a$ be $\psi$, then $x-a$ is $\frac{\sqrt{\gamma_i}z}{\sqrt{1-\gamma_i}}$. We know that for Taylor expansion, $x-a$ has to be a very marginal value. $\frac{\sqrt{\gamma_i}z}{\sqrt{1-\gamma_i}}$ satisfies this. From the example given in the Result Analysis Section, $\gamma$ is generally below 0.2 and $z$ follows a Gaussian or NIG distribution, ranging between $[-1,1]$ with a value close to zero at most times. Therefore, Taylor's theorem for this task is reasonable. 

In this scenario, function $g(x)$ is ${\rm sin}^{-1}\sqrt{F(x)}$. It is a composite function, so the chain rule should be considered for the derivative function $g'(a)$.  ${\rm sin}^{-1}(h)'$= $\frac{1}{\sqrt{1-h^2}}$ and $(\sqrt h)'$ = $-\frac{1}{2\sqrt{h}}$ apply for any variable $h$, and note that $F$ is the cumulative distribution function (CDF), the derivative of $F$ would just be the probability density function (pdf) which is normally denoted as $f$. Now $g'(a)$ reads:
\begin{equation}
g'(a)=\frac{-f(\psi)}{2\sqrt{1-F(\psi)}\sqrt{F(\psi)}}
\end{equation}

Therefore we get:
\begin{equation}
\begin{split}
{\rm sin}^{-1}\sqrt{F(\psi-\frac{\sqrt{\gamma_i}z}{\sqrt{1-\gamma_i}})}=\\{\rm sin}^{-1}\sqrt{F(\psi)}+\frac{-f(\psi)}{2\sqrt{1-F(\psi)}\sqrt{F(\psi)}}&\frac{\sqrt{\gamma_i}z}{\sqrt{1-\gamma_i}}
\end{split}
\end{equation}

One can now very easily get the expression for \emph{slope} and \emph{offset}:
\begin{equation}
\emph{slope} =\frac{-\sqrt{\gamma_i}}{2\sqrt{1-\gamma_i}} \frac{f(\psi)}{\sqrt{1-F(\psi)}\sqrt{F(\psi)}} 
\end{equation}
\begin{equation}
\emph{offset}=2 {\rm arcsin} (\sqrt{F(\psi)})
\end{equation}

Similarly, one can derive a second-order Taylor formula for this approximation:
\begin{equation}
g(x)=g(a)+g'(a)(x-a)+\frac{g''(a)}{2}(x-a)^2
\end{equation}
where $g''(a)$ is the second derivative. Knowing the expression for $g'(a)$ from Eq.(A14), we denote $D=2\sqrt{1-F(\psi)}\sqrt{F(\psi)}$, then derivative for $D$ is: 
\begin{equation}
\begin{split}
D'=(\frac{\sqrt{F(\psi)}}{\sqrt{1-F(\psi)}}+\frac{\sqrt{1-F(\psi)}}{\sqrt{F(\psi)}})&f(\psi)\\
=\frac{f(\psi)}{\sqrt{1-F(\psi)}\sqrt{F(\psi)}}
\end{split}
\end{equation}
Then let's derive $g''(a)$:
\begin{equation}
\begin{split}
g''(a)=(\frac{-f(\psi)}{D})'=\frac{-f'(\psi)D-D'(-f(\psi))}{D^2}\\
=\frac{-2f'(\psi)\sqrt{1-F(\psi)}\sqrt{F(\psi)}+\frac{f^2(\psi)}{\sqrt{1-F(\psi)}\sqrt{F(\psi)}}}{4(1-F(\psi))F(\psi)}
\end{split}
\end{equation}
So the second order term is:
\begin{equation}
\begin{split}
\frac{g''(a)}{2}(x-a)^2=\frac{\gamma_i[f^2(\psi)-2f'(\psi)(1-F(\psi))F(\psi)]}{8(1-\gamma_i)(1-F(\psi))^\frac{3}{2}F(\psi)^\frac{3}{2}}z^2
\end{split}
\end{equation}

The difference of using the first-order approximation in Eq.(A13) and the second-order approximation in Eq.(A18) is shown in Fig.A5a. The difference is very marginal, especially when most $Z$ values are located around zero following a Gaussian distribution. In the built-in code in Qiskit module, the first order Taylor's theorem is used. This can be enough to cope with the approximation. The circuits for first-order and second-order approximation are shown in Fig.A5b and A5c, respectively. If one wants to derive a second-order expansion, it involves much larger circuit depth, that is, a depth of 29 comparing to 5 for the first-order one, so we generally just consider the first order Taylor's theorem for this task. 

\begin{figure*}[t!]
\includegraphics[width=0.78\textwidth]{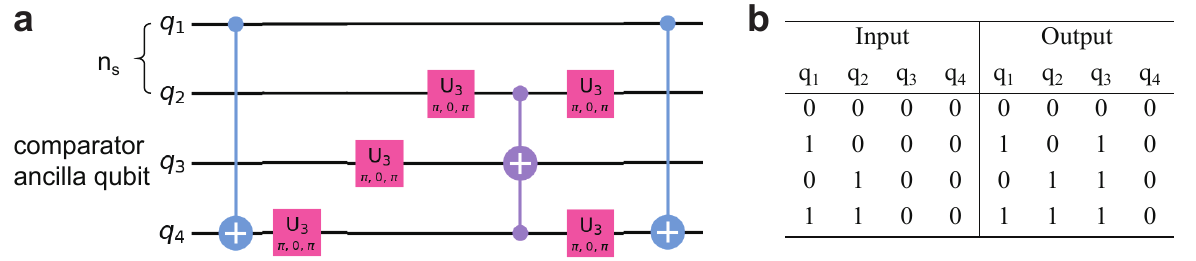}
\caption{\textbf{The comparator part.} (\textbf{a}) The quantum circuit for the comparator to compare with 1. The $U3(\pi, 0,\pi)$ gate is in fact an $X$ gate. (\textbf{b}) The truth table for this compartor part. The comparator ancilla qubit $q_3$ will be turned on when at least one of $q_2$ and $q_1$ is 1.}
\label{fig:assetdefaul}
\end{figure*}

\begin{figure*}[t!]
\includegraphics[width=0.98\textwidth]{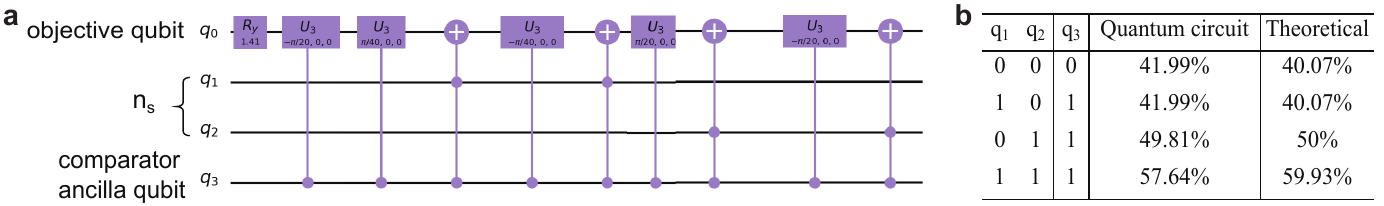}
\caption{\textbf{The linear rotation part.} (\textbf{a}) Quantum circuit for the initial linear rotation that loads $g_0$ and $g_z$. (\textbf{b}) The probability for the objective qubit at state $\ket 1$. `Quantum circuit" refers to the probability for the comparator ancilla qubit $q_3$ at state $\ket 1$ via matrix calculation for the quantum circuit shown in (\textbf{a}). `Theoretical' refers to the result calculated using Eq.(2) of the main text. }
\label{fig:assetdefaul}
\end{figure*}

~\\
\section*{Appendix VII Quantum circuit for the operator $\mathcal{S}$ }

As we have mentioned, the operator $\mathcal{S}$ requires $2n_s$ qubits, where $n_s={\rm floor}[{\rm log_2}(\sum_{i=1}^{n_x}\lambda_i)]+1$ ensures that the maximal sum of loss given default when all assets default can be encoded in $n_s$ qubits. As illustrated in Fig.A6a, among the $2n_s-1$ qubits, the first $n_s$ qubits are used for representing the calculated sum, and the following $n_s-1$ qubits are used as the carry qubits. For instance, when a binary number 011 is added by 1, then the sum becomes 100, and the least and second least digits have switched on a carry qubit to be 1.

The part that has been circled out in Fig.A6a is just one unit for the sum operator, and the sum operator totally includes three units for the three assets each with the same $\lambda$. If the $\lambda$s are in different integers, for instance, $\lambda_1=1$ and $\lambda_2=2$, then we can just repeat the unit twice for loading $\lambda_2=2$. In order to understand why such an operator can implement the sum operation, we show the circuit decomposition and in Fig.A6b truth table for how the circuit will output for all senarios. Note that the input of $q_{s1}$ and $q_{s2}$ can't both be 1, because we've ensured that even after adding 1 from $q_lz$ the sum would not exceed the value that $n_s=2$ qubits could convey. The component $S_a$ as marked in Fig.A6a decides whether to switch on the carry qubit or not. The component $S_b$ changes the lower and higher digit to load the sum. The component $S_c$ ensures the the carry qubit clears to $\ket 0$.

~\\
\section*{Appendix VIII Piecewise linear rotation for objective qubit }

We use the comparator operator $\mathcal{C}_{L_k}$ ($k$=1, 2 and 3) to compare the sum of loss with the fixed lower attachment point $K_{L_k}$ for each Tranche $k$. The comparator has been used to compare the underlying asset value with the striking price for option pricing in a recent work[Ref A3], where the detailed quantum comparator circuit has been given.

The operator $\mathcal{C}_{L_k}$ would flip a qubit from $\ket 0$ to $\ket 1$ if $L(z)$, the sum of loss under the systematic risk $Z$, is higher than $K_{L_k}$, and would keep $\ket 0$ otherwise. For simplicity, we just discuss $\mathcal{C}_L$ generally that can later apply to all $\mathcal{C}_{L_k}$s by just setting $K_L$ to $K_{L_k}$. Meanwhile, the objective qubit will also rotate its state under the control of the comparator ancilla qubit. 

As suggested by the output state in Eq.(8) of the main text, the probability at state $\ket 1$ that can be measured using QAE would be expressed as follows:
\begin{equation}
P_1=\sum_{L(z)\leq K_L} f(z){\rm sin}^2(g_0)+\sum_{L(z)>K_L} f(z){\rm sin}^2(g_0+g_z)
\end{equation}
where $f(z)$ shows the probability of a systematic risk value $z$ distributed following a certain probability density function $f$. 
Given that ${\rm sin}^2(x + \frac{\pi}{4}) = x + \frac{1}{2} + \mathcal{O}(x^3)$, so ${\rm sin}^2(g_0)={\rm sin}^2(\frac{\pi}{4}-c)=\frac{1}{2}-c-\mathcal{O}(c^3)$ and for marginal $c$: ${\rm sin}^2(g_0)=\frac{1}{2}-c$. Therefore, the expression for probability $P_1$ can be further derived:
\begin{equation}
\begin{split}
P_1=\sum_{L(z)\leq K_L} f(z)(\frac{1}{2}-c)+\\\sum_{L(z)>K_L} f(z)(\frac{1}{2}-&c+2c\frac{min(L(z), K_U)-K_L}{K_U-K_L})\\
=(\frac{1}{2}-c)+\sum_{L(z)>K_L} f(z&)(2c\frac{min(L(z), K_U)-K_L}{K_U-K_L})\\
=(\frac{1}{2}-c)+\frac{2c}{K_U-K_L}(&\mathbb{E}[L_{tranche}])
\end{split}
\end{equation}
where $\mathbb{E}[L_{tranche}]$ is the expectation of loss for a certain tranche, for instance, setting $K_L$ and $K_U$ to be $K_{L_1}$ and $K_{U_1}$, we get the loss for the Equity Tranche. Therefore, when we have obtained $P_1$ from the QAE circuit, we would be able to get the tranche loss and return.

\begin{figure*}[t!]
\includegraphics[width=0.98\textwidth]{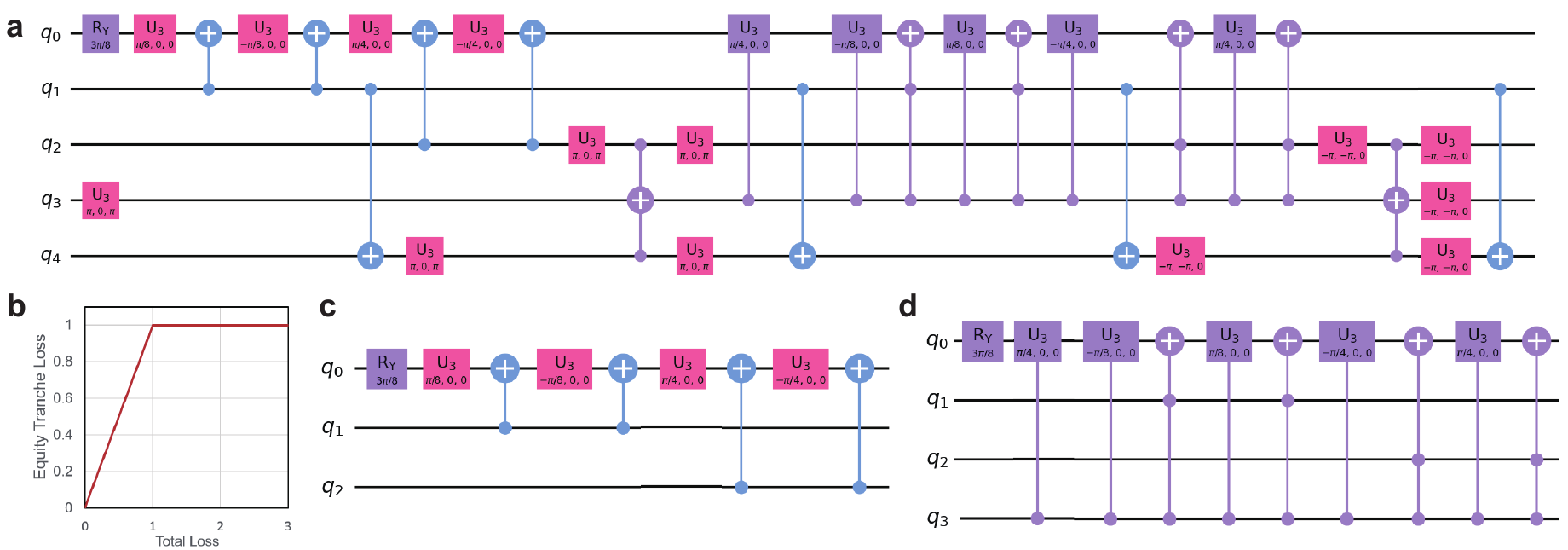}
\caption{\textbf{Quantum circuit for the operator $\mathcal{C\&R}$ for an equity tranche.} (\textbf{a}) The quantum circuit. Here $q_0$ is the objective qubit. $q_1$ and $q_2$ are the $n_s$ qubits that load the sum from the sum operator, and $q_3$ is the comparator ancilla qubit. (\textbf{b}) The payoff function for this equity tranche. (\textbf{c}) The first linear rotation part that's corresponding to the attachment point of 0. (\textbf{d}) The second rotation part that's corresponding to the attachment point of 1. }
\label{fig:assetdefaul}
\end{figure*}

\section*{Appendix IX Quantum circuit for the operator $\mathcal{C\&R}$ }
Before presenting the circuit for the whole operator $\mathcal{C\&R}$, we need to know that it includes several major components, including loading comparator, loading $g_0$ and $g_z$. We consider a very simple case, that is, the operator will rotate when the sum is above an attachment point of 1.

Firstly, the comparator circuit is presented in Fig.A7a, together with the truth table for this circuit. $q_2q_1$ can present four numbers: 00, 01, 10 and 11. As it is to compare with 1, the comparator ancilla qubit $q_3$ will be turned on when at least one of $q_2$ and $q_1$ is 1. The last part of the component ensures that $q_4$ always clears to $\ket 0$ after the comparison operation.

Then for the linear rotation gate, it is demonstrated in Fig.A8a. For $g_0={\rm sin}(\frac{\pi}{4}-c)$, it can simply be loaded by an $R_Y$ rotation gate, with an angle $2(\frac{\pi}{4}-c)$, so that the probability at $\ket 1$ becomes ${\rm sin}^2(g_0)$. Then the linear rotation, which involves comparator ancilla qubit $q_3$ and the objective qubit $q_0$, implement the rotation for $g_z$ under the control of the comparator qubit. We present in Fig.A8b the probability of objective qubit at state $\ket 1$ for different input sum values that the quantum circuit outputs via matrix calculation for quantum circuits, and they show consistent result from the theoretical calculation, that is, ${\rm sin}^2(g_0)$ when the sum is smaller than attachment point 1, and ${\rm sin}^2(g_0+g_z)$ when the sum is higher than 1.

In order to demonstrate how this operator works for the CDO tranche pricing with multiple attachment points, we demonstrate a full circuit for an comparator and linear rotation operator $\mathcal{C\&R}$ in Fig.A9a for pricing the equity tranche, the payoff function of which is shown in Fig.A9b. Here $q_0$ is the objective qubit. $q_1$ and $q_2$ are the $n_s$ qubits that load the sum from the sum operator. $q_3$ is the comparator ancilla qubit. This circuit includes four components, namely, loading comparator, loading $g_0$, loading $g_z$ and clearing the comparator ancilla qubit. It is not difficult to figure out the two rotation parts of this circuit, and we plot them separately in Fig.A9c and A9d. They correspond to two attachment points, 0 and 1. For the first rotation part, as it's comparing with 0, so we don't need it to be connected to any comparator ancilla qubit. For the second rotation part shown in Fig.A9c, it is quite similar to the one shown in Fig.A8a. Still worth to notice, here the rotation is to flatten the payoff function, so that the setting of U3 gates is to multiply a minus sign comparing to the settings in Fig.A8a. and Fig.A9c, and this creates a negative rotation angle that make the payoff function flattened. 

~\\
\section*{Appendix X Quantum Amplitude Estimation }

Given a Boolean function $f$: to find an $x \in X$ where $X \rightarrow \{ 0,1 \}$ such that $f \left( x \right) =1$, we can denote $N$ as the number of inputs on which $f$ takes the value 1, and it can be written as $N= \vert \{ x \in (X \vert f ( x ) =1 \} \vert$.
If we have a classical probabilistic algorithm $P$ that outputs a guess on input \emph{x}, the solution to instance $x$ can be found by repeatedly calling  $P$ and $X$. If $X \left( x,P \left( x \right)  \right) =1$  with probability  $p>0$, we have to repeat the process $\frac{1}{p}$ times on average.

\begin{figure*}[t!]
\includegraphics[width=0.97\textwidth]{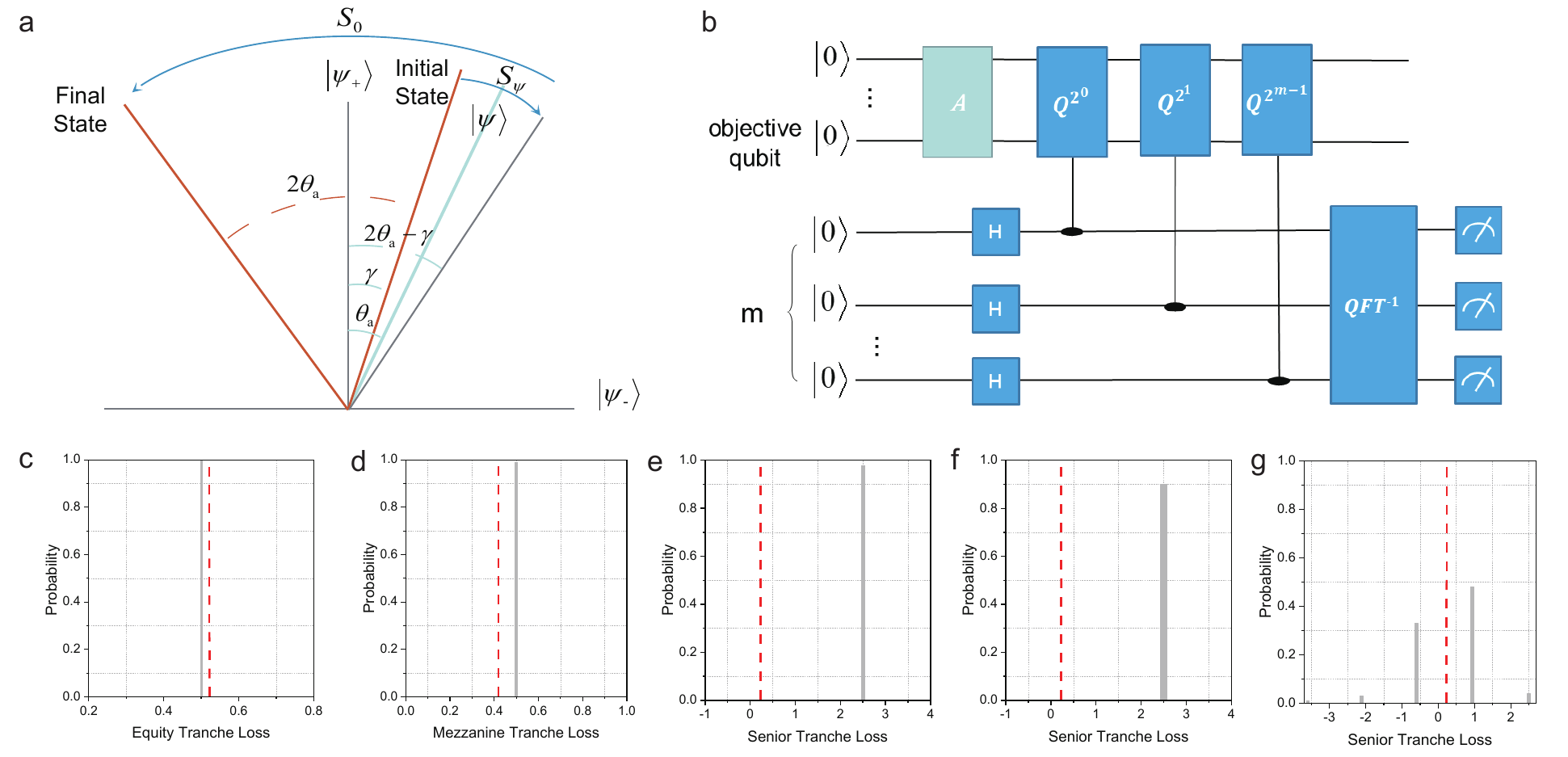}
\caption{\textbf{Theoretical framework for QAE.} (\textbf{a}) Illustration for angle rotation by operator $Q$. (\textbf{b}) The quantum circuit for quantum phase estimation. $H$ denotes the Hadamard gate. $QFT^{-1}$ denotes the inverse quantum Fourier transform. (\textbf{c-g}) The calculated tranche loss for (\textbf{c}) equity tranche using $m=$, (\textbf{d})mezzanine tranche using , (\textbf{e}) senior tranche using $m=2$, (\textbf{f})senior tranche using $m=3$, and (\textbf{g}) senior tranche using $m=7$. In (\textbf{c-g}), the grey bars indicate the canonical QAE results and the red dashed lines indicate the matrix calculation results for the quantum circuits.
 }
\label{fig:CDOTrancheStructure}
\end{figure*}

\begin{figure*}[t!]
\includegraphics[width=0.98\textwidth]{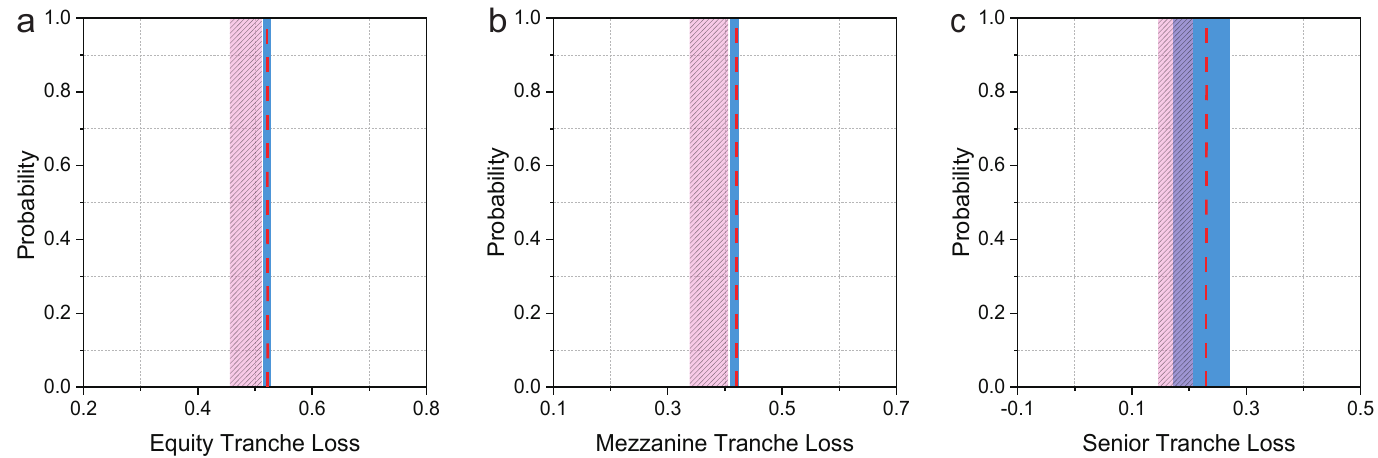}
\caption{\textbf{CDO tranche loss with $Z$ under the Gaussian distribution.} The calculated loss for (\textbf{a}) the Equity Tranche, (\textbf{b}) the Mezzanine Tranche, and (\textbf{c}) the Senior Tranche. The systematic risk $Z$ follows the Gaussian distribution (mean=0, variance=1). Blue bars indicate quantum computation results with $\epsilon=0.002$ and $\alpha=0.05$ for IQAE using QASM simulator. Red dashed lines indicate definite matrix calculation result of the quantum circuit. The pink shading areas indicate the Monte Carlo results, which are obtained by finding the range of 20 sets of Monte-Carlo simulations, each set having 1000 random repetitions. }
\label{fig:CDOTrancheStructure}
\end{figure*}

Suppose given a unitary transformation, $\mathcal{A}$, which is a quantum algorithm, unlike making measurement in the case of classical $P$ algorithm, this produces a quantum superposition state of the $``$desired$"$ result that $X \left( x,P \left( x \right)  \right) =1$ and $``$undesired$"$  result that $X \left( x,P \left( x \right)  \right)  \neq 1$. Then amplitude estimation is the problem of estimating $a$, the probability that a measurement of $\ket \psi$ yields a good solution. It is sufficient to evaluate $\mathcal{A}$ and $X$ in an expected number of times that is proportional to  $ \frac{1}{\sqrt[]{a}}~. $ \par

To explain amplitude estimation, the quantum state after unitary transformation $\mathcal{A}$ can be expressed as a linear combination of $\psi _{+}$ and the orthogonal $\psi _{-}$:
\begin{equation}
 \mathcal{A} \vert 0 \rangle = \vert \psi\rangle = -\frac{\emph{i}}{\sqrt[]{2}}\left(e^{\emph{i} \theta _{a}}\vert \psi _{+} \rangle  + e^{-\emph{i} \theta _{a}}\vert \psi _{-} \rangle \right) 
\end{equation}
so the success probability $``a"$ is converted to the solution of angle $\theta _{a}$ that decides the eigenvalue for unitary transformation $\mathcal{A}$. 

Apply an operator $Q$ to $\mathcal{A}$:
\begin{equation}
Q= AS_0A^{\dagger} S_{\psi 0}
\end{equation}
where $S_0 =1-2\ket 0 \bra 0$, and $S_{\psi 0} =1-2\ket \psi \ket 0 \bra 0 \bra \psi$. As illustrated in Fig.A10a, an initial state first rotates along $\psi$ by the operator $S_{\psi 0}$, and then rotates along $\psi _{+}$ by the operator $S_0$. Therefore, the angle between the arbitrarily set initial state and the final state after the operator $Q$ becomes $2\theta _{a}$. In this case when the initial state is $\mathcal{A}$, operator $Q$ just shifts from $\ket \psi$ for $2\theta _{a}$.

The task of finding the eigenvalue for quantum state $\ket \psi$ of the unitary transformation $\mathcal{A}$ can be fulfilled by Quantum Phase Estimation that requires another register with $m$ additional qubits. As shown in Fig.A10b, the phase estimation quantum circuit comprises Hadamard gates, controlled-rotation operators and an inverse quantum Fourier transform ($\mathcal{QFT}^{-1}$) operation. 

Firstly, the Hamamard gates prepare the $m$ qubits in the uniform superposition:
\begin{equation}
\ket 0 ^{\otimes m} \ket \psi \rightarrow \frac{1}{\sqrt{2^m}}\sum_{j=0}^{2^m-1} \ket j \ket \psi
\end{equation}

As has been mentioned, the operator $Q$ essentially causes a Y-rotation of angle $2\theta _a$, $i.e.$, $Q=R_y(2\theta _a)$. In this phase estimation circuit, the many controlled-rotation operators $Q_j$ satisfies: $Q_j=R_y(2j\theta _a)$, and they turn the above equation into:
\begin{equation}
\frac{1}{\sqrt{2^m}}\sum_{j=0}^{2^m-1} e^{ 2i\theta_a j}\ket j  \ket \psi
\end{equation}

Applying the $\mathcal{QFT}^{-1}$ operation, we can reverse the action on vector $\ket j$ to that on $\ket {\theta_a}$:
 \begin{equation}
\mathcal{QFT}^{-1}\Big(\frac{1}{\sqrt{2^m}}\sum_{j=0}^{2^m-1} e^{ 2i\theta_a j}\ket j  \ket \psi \Big)=\frac{1}{\pi} \ket {\theta_a} \ket \psi
\end{equation}

By taking measurement on the register of $m$ qubits, we can get the approximation of $\theta_a$. This is done by obtaining the measured integer $y \rightarrow \left\{0,1,2, \ldots 2^m-1 \right \}$. Taking $M=2^m$, then $\theta_a$ can be approximated as $\widetilde{\theta_a}=y\pi/M$,  which yields $\widetilde{a}$, the approximation of the aforementioned probability $a$: 
\begin{equation}
\widetilde{a}=\sin ^{2} \left( \frac{y \pi }{M} \right) \in [0,1] 
\end{equation}
satisfying the following inequality:
\begin{equation}
| {a} - \widetilde{a} | \leq \frac{\pi }{M} + \frac{\pi^2 }{M^2} = O(M^{-1}) 
\end{equation} 
with probability at least  $\frac{8 }{M^2}$. Comparing with the $O(M^{-\frac{1}{2}})$ convergence rate of the classical Monte Carlo method, the quantum amplitude estimation method converges faster with a quadratic speed-up. 

In the content of this CDO tranche pricing task, the $\theta_a$ to be measured by QAE gives ${\rm sin}^2(\theta_a)$, which yields $P1$, the probability that contains the information of expected tranche loss as shown in Eq.(8). From Fig.A10c-g, we show QAE results for different tranches and using different $m$ values. It's suggested that for some tranche, such as the equity and mezzanine tranche, the estimation is relatively more accurate. However, for the senior tranche, as the expected tranche loss is very marginal, this adds difficulty for QAE calculation. Increasing $m$ which represents the precision of estimation from 2, 3, to 7, the QAE result using current Qiskit QASM simulator is relatively improving, but still not quite satisfatory. 

The method shown above is the canonical QAE method that has been raised since 2002. It yields quadratic speedup, but the use of the inverse Quantum Fourier Transform makes it require an exponentially increasing circuit depth, which is ineffficient. The iterative QAE (IQAE)[Ref A4] is one of the newly raised alternative QAE methods that would avoind those exponential consumption and is suggested to be more efficient. Same with canonical QAE, IQAE requires the rotation to make 
\begin{equation}
Q^k\ket \Psi={\rm sin}((2k+1)\theta_a)\ket 1+{\rm cos}((2k+1)\theta_a)\ket 0
\end{equation}
so the probability at such a state is ${\rm sin}^2((2k+1)\theta_a)$. 

Without using the inverse Quantum Fourier Transform, IQAE estimates the $\theta_a$ using the following method. We know that ${\rm sin}^2(x)=(1-{\rm cos}(2x))/2$, so the probability ${\rm sin}^2((2k+1)\theta_a)$ can be expressed as: 
\begin{equation}
(1-{\rm cos}((4k+2)\theta_a)/2
\end{equation}
Suppose the confidence interval $[\theta_u, \theta_l]$ for $\theta_a$, to find the largest $k$ such that $[(4k+2)\theta_u,(4k+2)\theta_l]_{mod 2\pi}$ is fully contained in the upper or lower plane. If we want to get $\tilde a =(a_l+a_u)/2$ as an estimator for $a$ with $|a-\tilde a|\leq \epsilon$ with a confidence of $1-\alpha$, we may need a number of shots up to $N_{max}(\epsilon, \alpha)$:
\begin{equation}
N_{max}(\epsilon, \alpha)=\frac{12}{{\rm sin^4} (\pi/30)}{\rm log}(\frac{2}{\alpha}{\rm log_3}(\frac{3\pi}{20\epsilon}))
\end{equation}

In practice, we input an initial interval $[\theta_u, \theta_l]$, and by increasing the value of $k$, we seek for  determines the largest feasible $k$ with $K=4k+2 \geq 2K_i$ such that $[K\theta_u, K\theta_l]_{mod 2\pi}$ lies either in the upper or lower plane. If the solution for such k exists, we can invert the cosine function and obtain an estimation for $\theta_{a}$.

\section*{Appendix XI Set input parameters for the built-in piecewise linear rotation function in Qiskit}
For the $\mathcal{C}\&\mathcal{R}$ part of the quantum circuit shown in Fig.2 of the main text, we can use the built-in code named `PwlObjective' for piecewise linear rotation function that includes the comparator $\mathcal{C}$, and the piecewise linear rotator $\mathcal{R}$. The built-in function uses the `breakpoints' array to record the attachment points, and uses the `slopes' and `offsets' arrays in which slope $k$ and offset $k$ correspond to these for the line segment between breakpoint $k-1$ and breakpoint $k$. Note the offset is the $y$-axis value for the starting point of the line segment, instead of the intercept by extending the line segment to the $y$ axis. The breakpoints, slopes and offsets for the tranche loss function are shown in each figure in Fig. 3c-e, which can be very straightforwardly calculated. These are used as the input parameters for the built-in piecewise linear rotation function. 

\begin{figure*}[hbt!]
\includegraphics[width=1.02\textwidth]{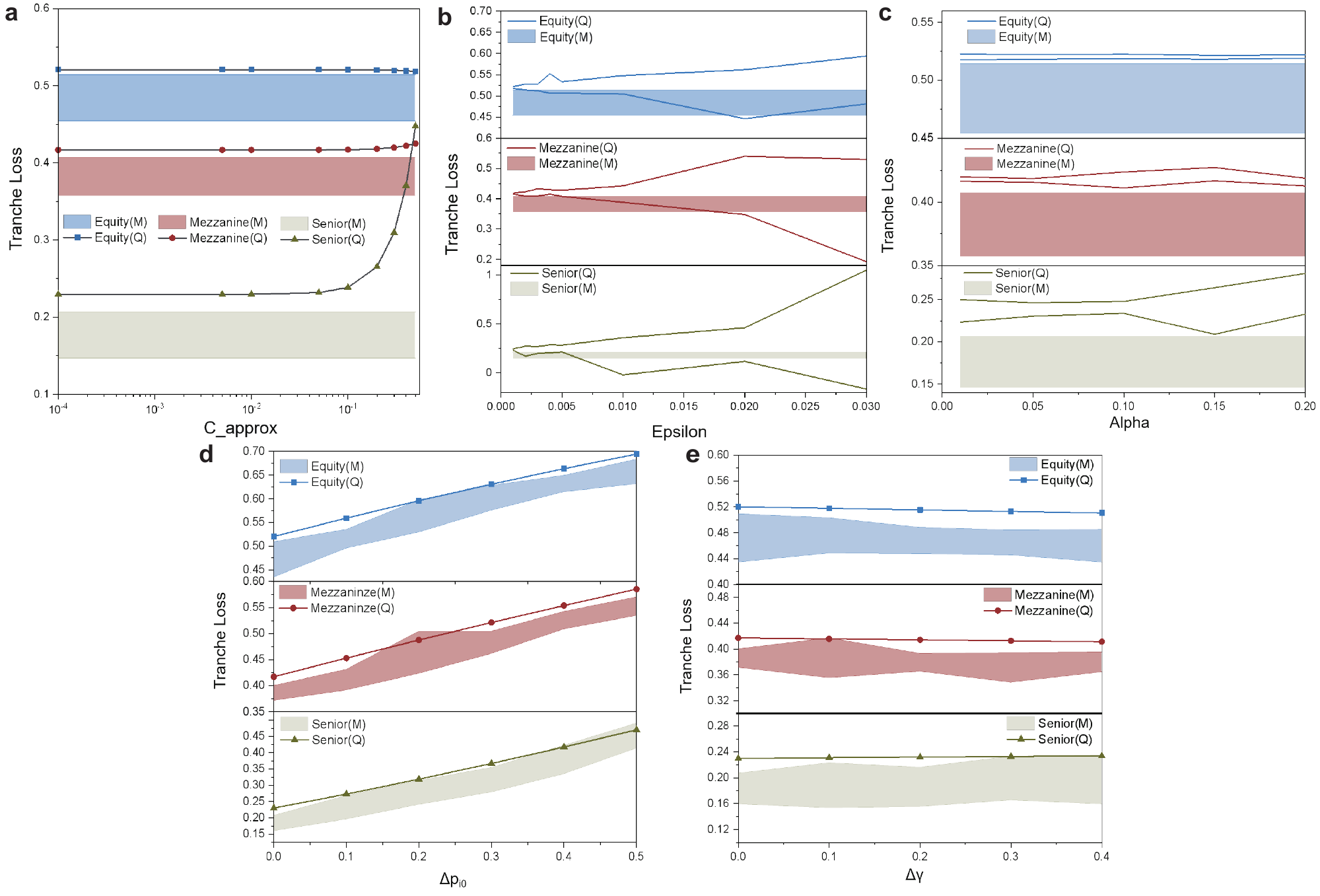}
\caption{\textbf{The robustness and accuracy for the tranche loss result.} (\textbf{a}) The calculated tranche loss via matrix calculation of quantum circuits when varing the scaling factor $c$ in the operator $\mathcal{C\&R}$. (\textbf{b-c}) The analysis of IQAE parameters $\epsilon$ and $\alpha$. The calculated tranche loss via running IQAE in QASM simulator when changing (\textbf{b}) $\epsilon$ while fixing $\alpha$ to be 0.05, and (\textbf{c}) changing $\alpha$ while fixing $\epsilon$ to be 0.001. (\textbf{d-e}) The calculated tranche loss via matrix calculation of quantum circuits when (\textbf{d}) changing the input parameters $p_i^0$s while fixing original $\gamma_i$s, and (\textbf{e}) changing $\gamma_i$s while fixing original $p_i^0$s.  In (\textbf{a-e}),  (Q) stands for the quantum computation result, and (M) stands for the Monte-Carlo results are obtained by finding the range of 20 sets of Monte-Carlo simulations, each set having 1000 random repetitions. }
\label{fig:CDOTrancheStructure}
\end{figure*}

\begin{figure*}[hbt!]
\includegraphics[width=1.00\textwidth]{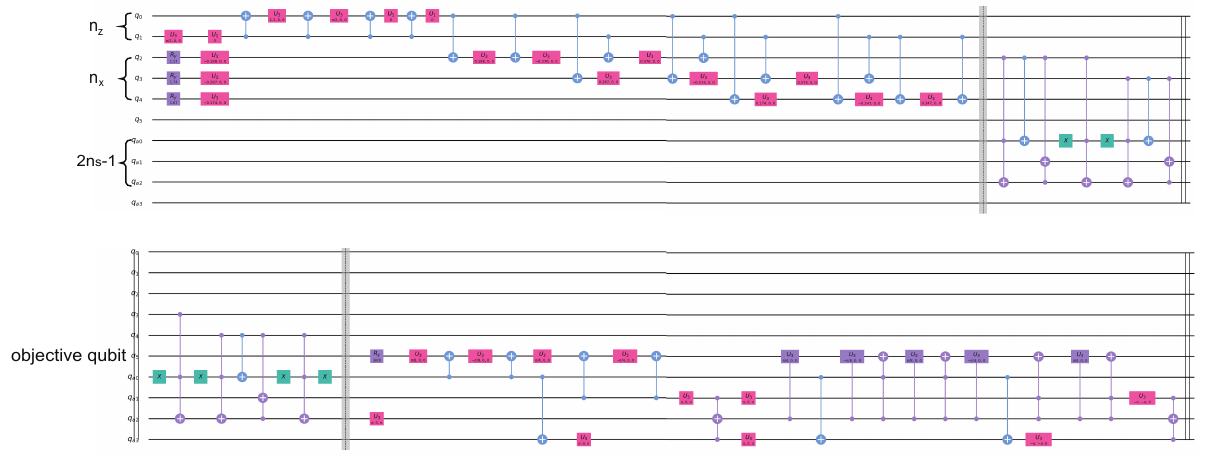}
\caption{\textbf{The full circuit for the equity tranche pricing in the simpler case.} The circuit before the first barrier is for the operator $\mathcal{L_X}$, $\mathcal{U_Z}$ and $\mathcal{L_Z}$. The circuit between the first and second barrier is for the operator $\mathcal{S}$, and the circuit after the second barrier is for the operator $\mathcal{C\&R}$. }
\label{fig:CDOTrancheStructure}
\end{figure*}

\section*{Appendix XII Discussion on tranche return in reality } 

It's worth noting that the returns for Equity and Mezzanine Tranche in the case study of the main text are a bit too high, comparing to the custom returns that would be around 15-25\% and 5-15\% for the Equity and Mezzanine Tranche, respectively[Ref A5]. It's partially because that default probabilities $p_i$s are a bit high. One more reason is that we ignore the recovery rate of the asset in order to focus on the essential structure. The recovery rate $\eta$, which is generally set as 40\%, means that when asset defaults, some values can be recovered by ways like selling real estates to get funds to compensate investors. Then the maximum loss would equal to the total notional value multiplies (1-$\eta$). In this example, the loss given default $\lambda_1$ to $\lambda_4$ would become 1.2, 1.2, 0.6 and 1.2, while the tranche attachment points keep unchanged. This would bring down the tranche loss.

\section*{Appendix XIII Discussion on the robustness and accuracy of the method}

As shown in Fig.A12, we have demonstrated the influence of a few parameters that are considered in quantum computation. We use the Monte Carlo simulation results as a benchmark, which, even if not very precise, can still show a range of reasonable results.  

We firstly consider the scaling factor $c$ that is set in the operator $\mathcal{C\&R}$. As discussed in Appendix IX, the approximation ${\rm sin}^2(g_0)=\frac{1}{2}-c$ stands only for a marginal value of $c$. Here we compare the expected tranche loss value via matrix calculation of quantum circuits when varying the scaling factor $c$, and we do find in Fig.A12a that a smaller scaling factor $c$ leads to a result closer to the Monte-Carlo calculation. Especially, the senior tranche loss result is more sensitive to $c$ than the other two tranches.

Then $\epsilon$ and $\alpha$ are two parameters considered in the iterative QAE module, as has been introduced in Appendix X. The iterative QAE does not give a definite value of amplitude estimation, but provides a confidence range of the result, and the final given estimated result is just the medium value of the range. Therefore, we show in Fig.A12b and A12c how the confidence range varys with the parameter $\epsilon$ and $\alpha$. The value of $\alpha$ that decides the confidence interval (which equals 1-$\alpha$ slightly influences on the range of result. On the other hand, the value of $\epsilon$ that decides the precision of estimation strongly influences the result range. At an $\epsilon$ of above 0.03, the range is too wide to make a precise estimation. How to find the optimal parameters for iterative QAE is an open question for the community right now and is worthy of further investigation.  

We also test how the calculated tranche loss would vary with the input asset parameters, the independent default probabilities $p_i^0$s and the correlation to the systematic risk $\gamma_i$s. The calculated tranche prices always increase with the increase of $p_i^0$s and slightly decreases of $\gamma_i$s, which are consistent with the Monte-Carlo simulation. 

\begin{figure}[hbt!]
\includegraphics[width=0.5\textwidth]{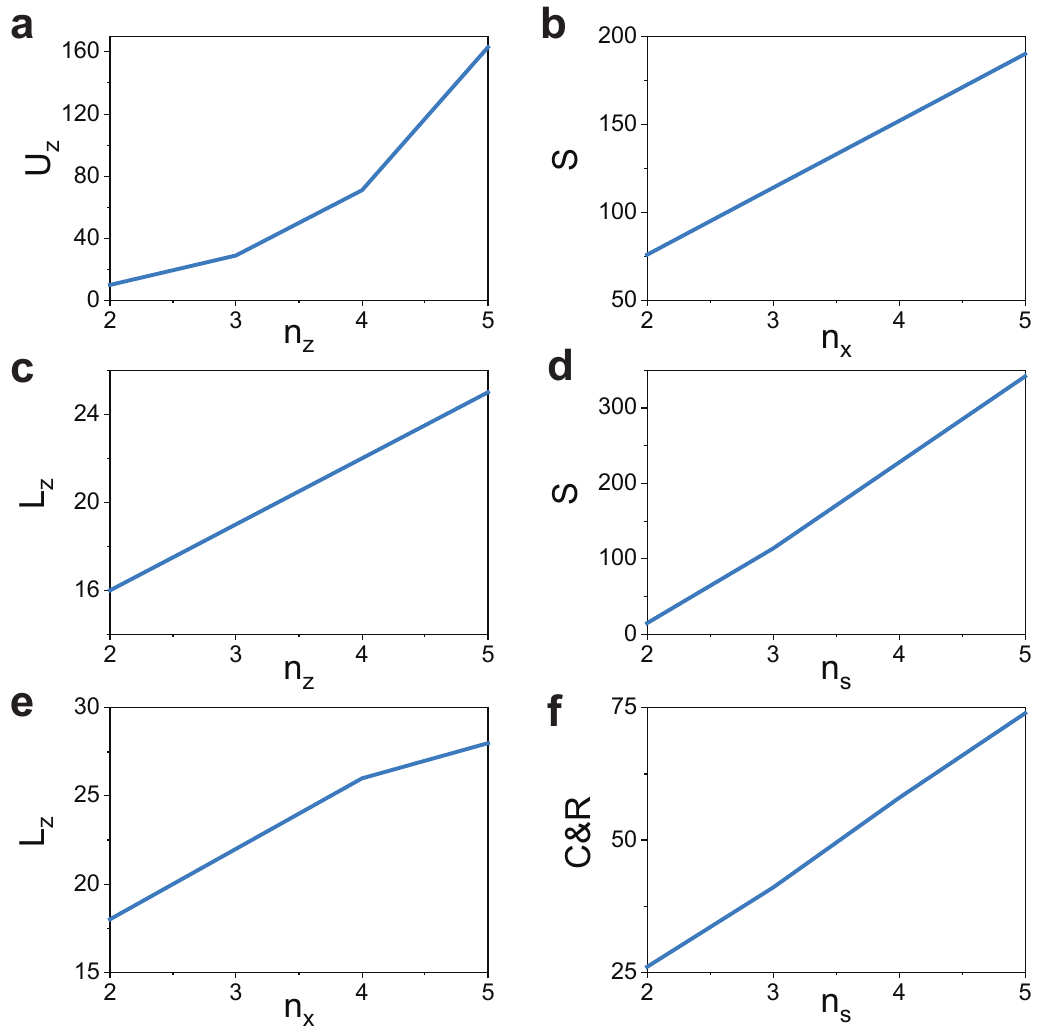}
\caption{\textbf{Scaling up the quantum circuit.} The circuit depth for (\textbf{a}) operator $\mathcal{U_Z}$ as a function $n_z$, (\textbf{b}) the operator $\mathcal{L_Z}$ as a function of $n_z$, (\textbf{c}) the operator $\mathcal{L_Z}$ as a function of $n_x$, (\textbf{d}) the operator $\mathcal{S}$ as a function of $n_x$, (\textbf{e}) the operator $\mathcal{S}$ as a function of $n_s$, and (\textbf{f}) the operator $\mathcal{C\&R}$ as a function of $n_s$.  }
\label{fig:CDOTrancheStructure}
\end{figure}

~\\
\section*{Appendix XIV Discussion on the scalability}
For the simple case that's analyzed in Appendix III-V, we present the full circuit for the equity tranche pricing task in Fig.A13, so that readers have a holistic view of the circuit. The function of each qubit has been marked in the circuit. The qubit $q_5$ is the objective qubit that loads the information $P1$, and the qubit $q_{a2}$ is used as the comparator ancilla qubit when it comes to the operator $\mathcal{C\&R}$ part. Before that, $q_{a2}$ is a carry qubit in the sum operator $\mathcal{S}$, and it has been cleared to $\ket 0$ (see Fig.A6) before it enters the $\mathcal{C\&R}$, so it is well prepared to be the initial state for a comparator ancilla qubit. 

For the pricing tasks for a larger scale, the quantum circuit cannot be easily plotted in one figure. For instance, for the task shown in the main text, we have provided the full circuit in the Supplementary Data. The circuit depth is often used to describe the size of a quantum circuit. In this task, the circuit depth for different components is generally related to $n_z$, the number of qubits used to load $Z$ distribution of systematic risk, $n_x$, the number of assets included in the asset pool, and $n_s$, the number of qubits required to load the maximal total sum. Fig.A14 shows the circuit dependence on those parameters for different components. 

For operator $\mathcal{L_X}$, it just includes one layer of $R_Y$ gates, so its depth is simply 1. We have seen in Fig.A4 that operator $\mathcal{L_z}$ can be divided into $n_x$ units, and each unit has controlled rotations that are controlled by each of the $n_z$ qubits. Therefore, we find operator $\mathcal{L_Z}$ almost increases linearly with $n_x$ and $n_z$ and the depth consumption is relatively economical. On the other hand, the circuit depth consumption for operator $\mathcal{U_Z}$ is increasing polynomially with $n_z$, taking the largest consuption among $\mathcal{L_X}$, $\mathcal{U_z}$ and $\mathcal{L_Z}$ when $n_z$ is four and above. Therefore, $\mathcal{U_Z}$ is the part we need to pay attention to if we want to limit the circuit depth.

For operator $\mathcal{S}$, the circuit depth roughly increases linearly with the number of assets $n_x$, and increases with $n_s$. For operator $\mathcal{C\&R}$, the circuit depth is firstly related to the number of attachment points it needs to compare with. We have shown in Fig.A9 that there are two linear rotation units corresponding to comparing with 0 and 1 respectively. Generally, the tranche pricing tasks require two attachment points for pricing each tranche. The circuit depth for $\mathcal{C\&R}$ is further related to $n_s$. A good thing is that, $n_s$ is the parameter we can control in practice. For instance, if three loss given defaults are 2,4 and 6, and the tranche attachment points are 4 and 8, then we can use scale the problem down by divide all loss given defaults and tranche attachment points by 2. The required $n_s$ would then reduces from 4 to 3. As for the scalability for QAE, it has been extensively analyzed in the recent works on improved QAE methods, that all trying to reduce the circuit depth to go below an exponentially increasing scale. This is a separate topic, and the investigation on it would be helpful for using it in wider applications.

~\\


\noindent [A1] Schl\"osser, A. Normal Inverse Gaussian Factor Copula Model. \emph{Normal Inverse Gaussian Factor Copula Model.} 80-93, Springer Berlin (2011).

\noindent [A2] Gu\'egan, D., \& Houdain, J. Collateralized Debt Obligations pricing and factor models: a new methodology using Normal Inverse Gaussian distributions. Note de Recherche IDHE-MORA No. 007-2005, ENS Cachan (2005).

\noindent [A3] Stamatopoulos, N., Egger, D. J., Sun, Y., Zoufal, C., Iten, R., Shen, N., \& Woerner, S. Option Pricing using Quantum Computers. \emph{Quantum} \textbf{4}, 291 (2019).

\noindent [A4] Grinko, D., Gacon, J., Zoufal, C., \& Woerner, S. Iterative Quantum Amplitude Estimation. \emph{arXiv Preprint}, arXiv:1912.05559 (2019).

\noindent [A5] Chacko, G., Sj\"oman, A., Motohashi, H., \& Dessain, V. \emph{Credit Derivatives, Revised Edition: A Primer on Credit Risk, Modeling, and Instruments.} Pearson Education (2016).

\end{appendix}

\end{document}